\documentclass[useAMS,usenatbib]{mn2e}
\usepackage[dvips]{graphicx}

 \title[Type Ib Supernova SN 2009jf]{Optical studies of SN 2009jf: A  type Ib supernova 
with an extremely slow decline and aspherical signature}

\author[D.K. Sahu et al.]
{D. K. Sahu$^1$, U.K. Gurugubelli$^{1,2}$, G. C. Anupama$^1$, K. Nomoto$^{3}$\\
1. Indian Institute of Astrophysics, Koramangala, Bangalore 560 034, India\\
2. Joint Astronomy Programme, Indian Institute of Science, Bangalore 560 012,
India\\
3. Institute for the Physics and Mathematics of the
Universe, University of Tokyo, Kashiwa, Chiba 277-8583, Japan \\
(E-mail: dks{@}iiap.res.in, uday{@}iiap.res.in, gca{@}iiap.res.in, nomoto{@}astron.s.u-tokyo.ac.jp )}

\begin{document}

\date{Accepted .....; Received ......}


\maketitle

\label{firstpage}

\begin{abstract}
Optical $UBVRI$ photometry and medium resolution spectroscopy of the type Ib 
supernova SN 2009jf, during the period $\sim -15$ to  $+250$ days with respect to the
$B$ maximum are reported. The light curves are broad, with an extremely slow decline. The 
early post-maximum decline rate in the $V$ band is similar to SN 2008D, however,
the late phase decline rate is slower than other studied type Ib supernovae. 
With an absolute magnitude of $M_{V} = -17.96\pm0.19$ magnitude at peak, 
SN 2009jf is a normally bright supernova.  The peak bolometric luminosity and 
the energy deposition rate via $^{56}$Ni $\rightarrow$ $^{56}$Co chain indicate 
that $\sim {0.17}^{+0.03}_{-0.03}$ M$_{\odot}$ of $^{56}$Ni was ejected during the 
explosion. He\,I 
5876~\AA\ line is clearly identified in the first spectrum of day $\sim -15$, at
a velocity of $\sim 16000$ km sec$^{-1}$. The [O\,I] 6300-6364~\AA\ line seen
in the nebular spectrum has a multi-peaked and asymmetric emission profile, with
the blue peak being stronger. The estimated flux in this line implies 
$\ga 1.34$ M$_\odot$ oxygen was ejected. The slow evolution of the light 
curves of SN 2009jf indicates the presence of a massive ejecta. The 
high expansion velocity in the early phase and broader emission lines during the
nebular phase suggest it to be an explosion with a large kinetic energy. A 
simple qualitative estimate leads to the ejecta mass of M$_{\rm ej} = 4-9$
M$_\odot$, and kinetic energy E$_{\rm K} = 3-8 \times 10^{51}$ erg. The 
ejected mass estimate is indicative of an initial main-sequence mass of 
$\ga 20- 25$ M$_\odot$. 

\end{abstract}


\section{Introduction}
Type Ib supernovae (SNe Ib) are core-collapse supernovae, characterized by the 
presence of prominent helium lines and the absence of hydrogen lines. They are 
believed to be the results of violent explosions of massive stars, such as the
Wolf-Rayet stars, which are stripped of most or all of their hydrogen envelope, 
either by mass transfer to a companion (e.g., \citealt{nomoto94}, \citealt{pods04}),
or via strong winds \citep[e.g.,][]{woosley93}, or by sudden eruptions. 
These supernovae are also termed as stripped-envelope supernovae.

The presence of hydrogen in type Ib supernovae remains  an open issue for 
investigation. There  are some  type Ib events which show a deep absorption 
at $\sim 6200$~\AA\ in their early spectra, which could be attributed to 
H$\alpha$ (\citealt{branch02}, \citealt{anupama05}, \citealt{soderberg08}), 
whereas some others show a shoulder in the red wing of the [O\,I] 6300-6364~\AA\
line in their nebular spectra, due to H$\alpha$ (\citealt{sollerman98}, \citealt
{strit09}). Using the SYNOW code, \cite{elmhamdi06} have shown the presence of a 
thin layer of hydrogen ejected at high velocity in almost all the SNe Ib in 
their sample. \cite{maurer10} have recently investigated various mechanisms
that can produce strong H$\alpha$ emission in the late phase, and shown that it 
can be explained well by radioactive energy deposition, if hydrogen and helium 
are mixed in suitable fractions and clumped strongly.  

Late phase observations of SNe Ib have gained special importance as these 
phases probe deeper into the core of the expanding stars. The nebular spectrum 
originating from an optically thin ejecta provides important clues to the 
nature of progenitor star and the explosion mechanism. Asphericity in the 
explosion of stripped envelope supernovae is confirmed by a higher degree of 
polarization through spectropolarimetric studies  of these objects during  early 
phases (\citealt{wang03}, \citealt{leonard06}).  Independent indications of 
the asphericity in the explosion come from the narrower width of [O\,I] 
6300-6364~\AA\ line compared to the [Fe\,II] features at $\sim 5200$~\AA\ 
(\citealt{mazzali01}, \citealt{maeda02}) and/or from the asymmetric profile of 
the [O\,I] 6300-6364~\AA\ line \citep{mazzali05}.    

SN 2009jf was discovered  by \cite{li09} in the Seyfert 2, barred spiral galaxy 
NGC 7479 on September 27.33. This supernova was classified as a young type Ib 
supernova by \cite{kasliwal09}, and \cite{sahu09a}, based on early spectra 
obtained on September 29. \cite{itagaki09} reported the detection of a dim 
object at an unfiltered  magnitude of $\sim 18.2$  in an image obtained on 
2006 November 8.499 and at a magnitude of $\sim 18.3$ in an image obtained on 
2007 August 13.74. They also report the presence of the object at $\sim 18$ 
magnitude in the DSSS images. They have estimated the absolute magnitude of the 
object as $-14.5$ and suggested that these may be recurring outbursts of a 
luminous blue variable. 

In this paper we report optical photometry and spectroscopy of SN 2009jf in the 
early and nebular phase and discuss the results based on the observations. 

\section{Observation and data reduction}
\subsection{Photometry}
Photometric observations of SN 2009jf began on 2009 September 29, using the 2m 
Himalayan Chandra Telescope (HCT) of the Indian Astronomical Observatory, 
immediately after discovery, and continued until 2010 June 21, with a break 
during the period the supernova was behind the Sun. The observations were made 
using the Himalaya Faint Object Spectrograph Camera (HFOSC). The central 
2K$\times$2K region of the 2K$\times$4K SITe CCD chip in use with the HFOSC was 
used for imaging. This provides an image scale of 0.296 arcsec pixel$^{-1}$ 
over a $10\times 10$~arcmin$^{2}$  field of view. Further details on the telescope and 
instrument can be obtained from ''http://www.iiap.res.in/centers/iao''. The 
supernova was observed in Bessell $U$, $B$, $V$, $R$ and $I$ filters. Standard 
fields PG0231+051, PG1657+078 and PG2213-006 \citep{landolt92} observed under 
photometric sky condition on 2009 September 30 and October 14, are used for  
photometric calibration of the  supernova magnitudes.

\begin{figure}
\resizebox{\hsize}{!}{\includegraphics{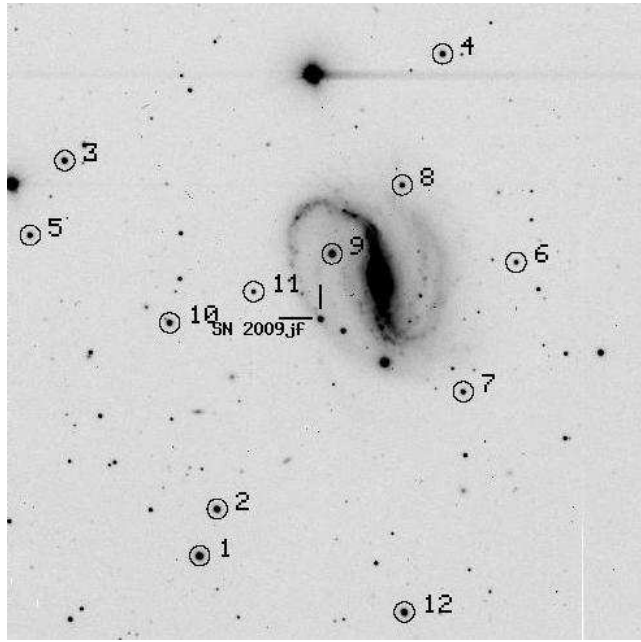}}
\caption[]{Identification chart for SN 2009jf. The stars used as local standards are
marked with numbers 1-12. South is up and east to the right. The field of view
is $10^\prime \times 10^\prime$.}
\label{fig_std}
\end{figure}

Data reduction was done in the standard manner, the data were bias subtracted, 
flat-fielded and cosmic ray hits removed,  using the standard tasks available
within the Image Reduction and Analysis  Facility (IRAF)  package. 
Aperture photometry was performed on the standard stars with an optimal 
aperture determined using the aperture growth curve method. Aperture correction 
between the optimal aperture and an aperture close to the full width half 
maximum (FWHM) of the stellar profile that had the maximum signal-to-noise 
ratio was determined using the brighter stars and then applied to the fainter 
ones. Average extinction values for the site \citep{stalin08} were used to 
correct for the atmospheric extinction and the average colour terms for the 
filter-detector system were used to get the photometric solutions, based on the
magnitudes of the stars in the standard fields. These were then used to 
calibrate a sequence of local standards in the supernova field observed on the 
same nights as the standard fields. The local standards were then used for the 
photometric calibration of the supernova magnitudes. The magnitudes of the local
standards in the supernova field are listed in Table \ref{tab_std} and the 
supernova field with the local standards marked is shown in Figure 
\ref{fig_std}. 

\begin{table*}
\caption{Magnitudes for the sequence of secondary standard stars in
the field of SN 2009jf.}
\begin{tabular}{lccccc}
\hline\hline
ID & U  & B & V &  R & I \\
\hline\hline
1 & $14.846\pm 0.063$ &  $14.717\pm 0.010$ & $14.052\pm 0.019$ & $13.639\pm 0.003$ & $13.218\pm 0.016$\\
2 & $16.260\pm 0.087$ &  $15.757\pm 0.006$&    $14.878\pm 0.015$ &   $14.340\pm 0.011$ & $13.834\pm 0.006$\\
3 & $16.197\pm 0.079$ &  $15.885\pm 0.004$&    $15.148\pm 0.024$ &   $14.739\pm 0.001$ & $14.317\pm 0.013$\\
4 & $16.366\pm 0.061$ &  $16.040\pm 0.005$&    $15.289\pm 0.011$ &   $14.849\pm 0.010$ & $14.421\pm 0.020$\\
5 & $16.228\pm 0.082$ &  $16.156\pm 0.006$&    $15.463\pm 0.023$ &   $15.061\pm 0.002$ & $14.620\pm 0.012$\\
6 & $19.942\pm 0.181$ &  $18.271\pm 0.010$&    $16.915\pm 0.014$ &   $16.071\pm 0.005$ & $15.320\pm 0.013$\\
7 & $18.775\pm 0.096$ &  $17.387\pm 0.001$&    $16.203\pm 0.017$ &   $15.475\pm 0.003$ & $14.860\pm 0.016$\\
8 & $17.128\pm 0.077$ &  $16.729\pm 0.004$&    $15.939\pm 0.022$ &   $15.455\pm 0.000$ & $14.996\pm 0.021$\\
9 & $16.054\pm 0.059$ &  $15.350\pm 0.021$&    $14.464\pm 0.030$ &   $13.958\pm 0.000$ & $13.519\pm 0.019$\\
10& $16.446\pm 0.057$ &  $16.088\pm 0.005$&    $15.266\pm 0.024$ &   $14.764\pm 0.004$ & $14.264\pm 0.017$\\
11& $17.209\pm 0.094$ &  $17.182\pm 0.007$&    $16.555\pm 0.015$ &   $16.139\pm 0.009$ & $15.720\pm 0.012$\\
12& $15.440\pm 0.065$ &  $15.332\pm 0.010$&    $14.659\pm 0.020$ &   $14.232\pm 0.005$ & $13.791\pm 0.012$\\
\hline
\end{tabular}
\label{tab_std}
\end{table*}
The magnitudes of the supernova and the secondary standards were estimated 
using point-spread function photometry, with a fitting radius equal to the FWHM 
of the stellar profile. The difference between the aperture and profile fitting 
photometry was estimated using bright standards in the field and applied to the 
supernova magnitude. The night-to-night zero points were estimated using the 
local standards in the supernova field and the supernova magnitudes calibrated 
differentially with respect to the local standards. The supernova magnitude in 
$U$, $B$, $V$, $R$ and $I$ bands are given in Table \ref{tab_snmag}.

\begin{table*}
\caption{Photometric observations of SN 2009jf}
\begin{tabular}{lccccccc}
\hline\hline
 Date & J.D. & Phase\rlap{*} & U & B & V & R & I\\
     & 2454000+ & (days)    &   &   &   &   &\\
\hline\hline

29/09/2009& 2455104.171 & $-15.29$ & $17.557\pm 0.030$ & $17.622\pm 0.020$ & $17.047\pm 0.019$ & $16.733\pm 0.024$& $16.578\pm 0.027$\\
30/09/2009& 2455105.173 & $-14.29$ & $17.168\pm 0.067$ & $17.275\pm 0.026$ & $16.739\pm 0.023$ & $16.443\pm 0.016$& $16.317\pm 0.016$\\
01/10/2009& 2455106.104 &  $-13.36$ & $16.945\pm 0.033$ & $16.990\pm 0.037$ & $16.508\pm 0.025$ & $16.219\pm 0.026$& $16.077\pm 0.019$\\
03/10/2009& 2455108.261 & $-11.20$ & & $16.441\pm 0.032$ & $16.001\pm 0.038$ & $15.774\pm 0.031$& $15.594\pm 0.032$\\
04/10/2009& 2455109.194 & $-10.27$ & $16.078\pm 0.040$ & $16.275\pm 0.035$ & $15.842\pm 0.034$ & $15.611\pm 0.032$& $15.458\pm 0.028$\\
08/10/2009& 2455113.119 &  $-6.34$ & $15.572\pm 0.043$ & $15.752\pm 0.032$ & $15.365\pm 0.012$ & $15.159\pm 0.013$& $14.980\pm 0.026$\\
14/10/2009& 2455119.308 & $-0.15$ & $15.582\pm 0.042$ & $15.594\pm 0.021$ & $15.092\pm 0.011$ & $14.857\pm 0.015$& $14.643\pm 0.010$\\
15/10/2009& 2455120.120 & $+0.66$ & $15.517\pm 0.028$ & $15.575\pm 0.025$ & $15.078\pm 0.019$ & $14.826\pm 0.024$& $14.638\pm 0.020$\\
16/10/2009& 2455121.313 & $+1.85$ & $15.580\pm 0.044$ & $15.618\pm 0.027$ & $15.082\pm 0.014$ & $14.816\pm 0.015$& $14.600\pm 0.037$\\
22/10/2009& 2455127.171 & $+7.71$ & $15.992\pm 0.035$ & $15.855\pm 0.011$ & $15.136\pm 0.009$ & $14.832\pm 0.020$& $14.588\pm 0.025$\\
24/10/2009& 2455129.265 & $+9.81$ & $16.153\pm 0.030$ & $15.988\pm 0.016$ & $15.186\pm 0.020$ & $14.864\pm 0.023$& $14.597\pm 0.025$\\
27/10/2009& 2455132.288 & $+12.83$ & $16.574\pm 0.039$ & $16.277\pm 0.028$ & $15.318\pm 0.024$ & $14.950\pm 0.034$& $14.649\pm 0.038$\\
31/10/2009& 2455136.187 & $+16.73$ &                      & $16.687\pm 0.027$ & $15.604\pm 0.029$ & $15.132\pm 0.024$& $14.777\pm 0.021$\\
06/11/2009& 2455142.047 & $+22.59$&                      & $17.118\pm 0.018$ & $15.933\pm 0.021$ & $15.403\pm 0.021$& $15.024\pm 0.044$\\
10/11/2009& 2455146.146 & $+26.69$&   $17.534\pm 0.056$ & $17.291\pm 0.018$ & $16.108\pm 0.019$ & $15.565\pm 0.013$& $15.138\pm 0.043$\\
14/11/2009& 2455150.057 & $+30.60$ &   $17.598\pm 0.087$ & $17.421\pm 0.031$ & $16.268\pm 0.012$ & $15.720\pm 0.020$& $15.246\pm 0.014$\\
17/11/2009& 2455153.156 & $+33.70$ &     &                  $17.454\pm 0.013$ & $16.353\pm 0.022$ & $15.823\pm 0.017$& $15.328\pm 0.029$\\
21/11/2009& 2455157.21 & $+37.75$ &     & $17.547\pm 0.020$ & $16.455\pm 0.017$ & $15.946\pm 0.021$& $15.439\pm 0.019$\\
25/11/2009& 2455161.048 & $+41.59$&     &                  $17.603\pm 0.025$ & $16.510\pm 0.032$ & $16.026\pm 0.031$& $15.537\pm 0.034$\\
30/11/2009& 2455166.188 & $+46.73$ &     &                  $17.647\pm 0.032$ & $16.594\pm 0.034$ & $16.116\pm 0.024$& $15.596\pm 0.017$\\
03/12/2009& 2455169.030 & $+49.57$ &     &                  $17.654\pm 0.031$ & $16.624\pm 0.014$ & $16.173\pm 0.032$& $15.653\pm 0.031$\\
18/12/2009& 2455184.040 & $+64.58$&     &                  $17.751\pm 0.020$ & $16.815\pm 0.020$ & $16.393\pm 0.016$& $15.855\pm 0.023$\\
23/12/2009& 2455189.036 & $+69.58$&     &                  $17.812\pm 0.033$ & $16.890\pm 0.017$ & $16.478\pm 0.025$& $15.964\pm 0.029$\\
29/12/2009& 2455195.151 & $+75.69$&     &                  $17.847\pm 0.024$ & $16.947\pm 0.028$ & $16.541\pm 0.012$& $16.061\pm 0.045$\\
09/01/2010& 2455206.041 & $+86.58$ &     &                  $17.887\pm 0.022$ & $17.079\pm 0.021$ & $16.692\pm 0.022$& $16.213\pm 0.043$\\
20/01/2010& 2455217.073 & $+97.61$ &     &                  $18.003\pm 0.020$ & $17.203\pm 0.019$ & $16.819\pm 0.016$& $16.326\pm 0.016$\\
27/01/2010& 2455224.073 & $+104.61$ &     &                  $18.151\pm 0.040$ & $17.295\pm 0.036$ & $16.946\pm 0.012$& $16.464\pm 0.020$\\
01/02/2010& 2455229.071 & $+109.61$ &     &                  $18.181\pm 0.049$ & $17.410\pm 0.019$ & $17.011\pm 0.014$& $16.563\pm 0.016$\\
01/05/2010& 2455318.442 & $+198.98$&     &                                    & $18.524\pm 0.038$ & $17.979\pm 0.025$& $17.785\pm0.033$\\
24/05/2010& 2455341.418 & $+221.96$ &     &                  $19.212\pm 0.046$ & $18.753\pm 0.024$ & $18.126\pm 0.022$& $18.013\pm 0.039$\\
26/05/2010& 2455343.406 & $+223.95$&     &                  $19.296\pm 0.029$ & $18.808\pm 0.031$ & $18.177\pm 0.052$& $18.019\pm 0.051$ \\
11/06/2010& 2455359.430 & $+239.97$&     &                  $19.456\pm 0.034$ & $19.075\pm 0.032$ & $18.427\pm 0.036$& $18.336\pm 0.038$ \\
21/06/2010& 2455369.353 & $+249.89$ &     &                  $19.462\pm 0.028$ & $19.109\pm 0.026$ & $18.539\pm 0.026$& $18.373\pm 0.034$\\
\hline
\multicolumn{8}{l}{\rlap{*}\ Observed phase with respect to the epoch of 
maximum in the $B$ band (JD 2455119.46).}
\end{tabular}			    
\label{tab_snmag}	    
\end{table*}			    

\subsection{Spectroscopy}
Spectroscopic observations of SN 2009jf were obtained during 2009 September 29 
(JD 2455104.16) and 2010 June 22 (JD 2455370.38). The spectra were obtained in 
the wavelength ranges 3500--7800~\AA\ and 5200--9250~\AA\ using grisms Gr\#7 and 
Gr\#8 available with HFOSC. The log of spectroscopic observations is given in 
Table \ref{tab_spec}. Arc lamp spectra of FeNe and FeAr were obtained for 
wavelength calibration. Spectroscopic data reduction was performed in the 
standard manner. All spectra were bias subtracted, flat-fielded and the one 
dimensional spectra extracted using the optimal extraction method \citep{horne86}.
Wavelength calibration was effected using the arc lamp spectra. The accuracy of 
wavelength calibration was checked using the night sky emission lines, and 
whenever necessary small shifts were applied to the observed spectra. The 
spectra were flux calibrated by correcting for the instrumental response using 
response curves estimated from the spectra of spectrophotometric standards that 
were observed on the same night. For the nights that standard star spectra were 
not available, the response curve obtained during observations on nearby nights 
were used. The flux calibrated spectra in the two regions were then combined, 
scaled to a weighted mean to give the final spectrum. This  spectrum was then 
brought to an absolute flux scale using zero points determined from broad-band 
$UBVRI$ magnitudes. The supernova spectra were corrected for the host galaxy 
redshift of $z=0.007942$ and dereddened for a total reddening of $E(B-V) = 0.112$, 
as estimated in Section 4. The telluric lines have not been removed from the spectra.

\begin{table*}
\caption{Log of spectroscopic observations of SN 2009jf.}
\begin{tabular}{lccc}
\hline\hline
Date & J.D. & Phase & Range \\
     & 2450000+ & days & \AA\  \\
\hline\hline
29/09/2009 & 5104.16  &    -15.30    &   3500-7800; 5200-9250\\ 
30/09/2009 & 5105.13  &    -14.33   &   3500-7800; 5200-9250\\
01/10/2009 & 5106.17  &    -13.29   &   3500-7800; 5200-9250\\ 
07/10/2009 & 5112.09  &     -7.37   &   3500-7800; 5200-9250\\
08/10/2009 & 5113.07  &     -6.39   &   3500-7800; 5200-9250\\ 
15/10/2009 & 5120.17  &     +0.71   &   3500-7800; 5200-9250\\
24/10/2009 & 5129.22  &     +9.76   &   3500-7800; 5200-9250\\
10/11/2009 & 5146.19  &     +26.73   &   3500-7800; 5200-9250\\
17/11/2009 & 5153.17  &     +33.71   &   3500-7800; 5200-9250\\
03/12/2009 & 5169.17  &     +49.71   &   3500-7800; 5200-9250\\
23/12/2009 & 5189.06  &     +69.60   &   3500-7800; 5200-9250\\
08/01/2010 & 5205.04  &     +85.58   &   3500-7800; 5200-9250\\
21/01/2010 & 5218.07  &     +98.61   &   3500-7800; 5200-9250\\
31/05/2010 & 5348.40  &     +228.94  &   5200-9250\\
22/06/2010 & 5370.38  &     +250.92  &   5200-9250\\

\hline
\end{tabular}
\label{tab_spec}
\end{table*}

\section{Optical light curves and colour curves}

\begin{figure}
\resizebox{\hsize}{!}{\includegraphics{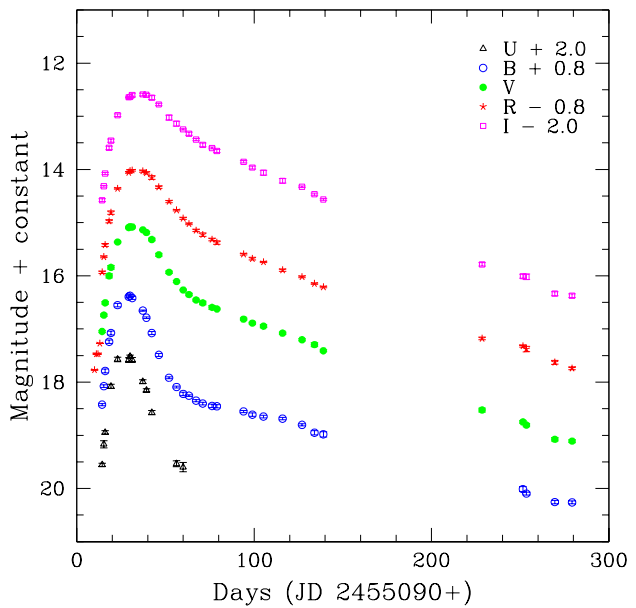}}
\caption[]{$UBVRI$ light curves of SN 2009jf. The light curves have been shifted by the 
amount indicated in the legend.}
\label{fig_light}
\end{figure}

The light curves of SN 2009jf in the $UBVRI$ bands are presented in Figure 
\ref{fig_light}. Also included in the figure are the unfiltered discovery 
magnitude and the pre-discovery limiting magnitudes \citep{li09}. Our 
observations began two days after discovery, $\sim$ 15 days before maximum 
in $B$ band and continued 
till $\sim 250$ days after maximum. The date of maximum brightness and the 
peak magnitude in  different bands have been estimated by fitting a cubic 
spline to the points around maximum and are listed in Table \ref{tab_peak}. 
Similar to other type Ib/c supernovae, the light curve of SN 2009jf peaks 
early in the bluer bands than the redder bands. The maximum in $B$ occurred on 
JD 2455119.46 (2009 October 14.9), at an apparent magnitude of $15.56\pm 0.02$. 

\begin{figure}
\resizebox{\hsize}{!}{\includegraphics{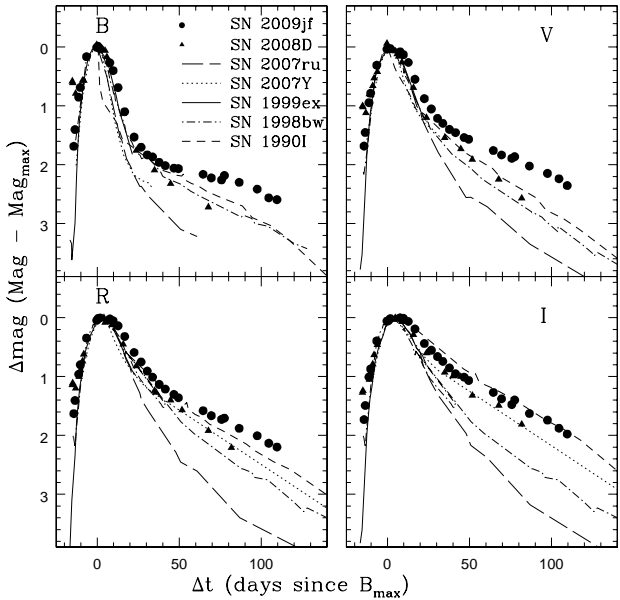}}
\caption[]{Comparison of $UBVRI$ light curves of SN 2009jf with those of 
SN 2008D, SN 2007Y, SN 1999ex, SN 1990I, SN 1998bw and SN 2007ru. The light curves have 
been normalized as described in the text. }
\label{fig_lcomp}
\end{figure}

The light curves of SN 2009jf are compared with those of a few well studied 
core-collapse supernovae, namely, SN 2008D \citep{modjaz09}, SN 2007Y 
\citep{strit09}, SN 1999ex \citep{strit02}, SN 1990I 
\citep{elmhamdi04}, broad lined type Ic SN 1998bw  and the fast declining broad-line 
type Ic SN 2007ru 
\citep{sahu09b}, in Figure \ref{fig_lcomp}. The observed magnitudes of the 
supernovae have been normalized to their respective peak magnitudes and shifted 
in time to the epoch of maximum brightness in $B$ band. From the figure, it is 
seen that the initial rise to maximum and post-maximum  decline of SN 2009jf is 
slower in comparison with other supernovae, making the light curve of SN 2009jf 
broader. The  post-maximum decline of the light curve in 15 days {\it i.e.} 
$\Delta m_{15}$ estimated for different bands are $\Delta m_{15}(U)= 0.971$,  
$\Delta m_{15}(B)=0.908$, $\Delta m_{15}(V)=0.503$, $\Delta m_{15}(R)=0.311$
and $\Delta m_{15}(I)=0.303$. These estimates indicate the decline rate in $V$ 
to be similar to that of SN 2008D, but considerably slower than other type Ib and type
Ic supernovae. 
The decline rates estimated for the later phases ($>40$ days) are 0.008 mag
day$^{-1}$ in $B$,  0.0126 mag day$^{-1}$ in $V$, 0.0141 mag day$^{-1}$ in $R$ 
and 0.0145 mag day$^{-1}$ in $I$. 
These rates are slower in comparison with other type Ib supernovae. The broader 
light curves and slower decline rates suggest that the ejecta of SN 2009jf is 
relatively efficient in trapping the $\gamma$-rays produced in the radioactive 
decay. This also indicates that the progenitor of SN 2009jf was able to retain 
more of an envelope prior to the core-collapse, thus increasing the diffusion 
time for the energy produced from the radio active decay of $^{56}$Ni to 
$^{56}$Co (\citealt{strit02},  \citealt{arnett82}).

\begin{table*}
\caption{Light curve maximum parameters}
\begin{tabular}{llll}
\hline\hline
 Filter  & J.D. at max. & Peak obs mag. & Peak abs mag. \\
     & 2455000+ &     &  \\
\hline\hline
U & 116.60$\pm$0.65 & 15.458$\pm$0.014 & -17.759$\pm$0.19\\ 
B & 119.46$\pm$0.50 & 15.558$\pm$0.020 & -17.579$\pm$0.19\\
V & 120.98$\pm$0.48 & 15.056$\pm$0.022 & -17.964$\pm$0.19 \\
R & 121.44$\pm$1.03 & 14.813$\pm$0.032 & -18.121$\pm$0.19\\
I & 124.10$\pm$0.60 & 14.586$\pm$0.010 & -18.253$\pm$0.19\\

\hline				    
\end{tabular}			    
\label{tab_peak}
\end{table*}			    
 
There are only a few cases of type Ib supernovae where the rise time to $B$ 
band maximum is constrained accurately. For example, the rise time for 
SN 1999ex and SN 2008D is found to be 18 days \citep{strit02} and 16.8 days 
\citep{modjaz09}, respectively. The rise time could be constrained for these 
two supernovae as they occurred in galaxies which were already being monitored 
to follow up other events. SN 1999ex was detected in IC 5179 which hosted 
SN 1999ee \citep{martin99}. The data obtained as a part of monitoring of 
SN 1999ee had the possible detection of the initial shock breakout due to SN 1999ex.  
Similarly, the shock  breakout of SN 2008D was detected by the {\it Swift} 
satellite, as an X-ray transient XRT 080109, during a routine follow up 
observation of SN 2007uy in NGC 2770 \citep{berger08}. The peak of the
X-ray transient is expected to occur shortly after the supernova explosion 
\citep{li07}. SN 2009jf was discovered on September 27.33, around 17.5 days 
before maximum in $B$ band. Comparing with SN 1999ex and SN 2008D, it appears
that SN 2009jf was most likely discovered almost immediately after explosion. 
However, since we are unable to constrain the shock breakout precisely, a rise
time of 19$\pm$1 days is assumed in this work. 

\begin{figure}
\resizebox{\hsize}{!}{\includegraphics{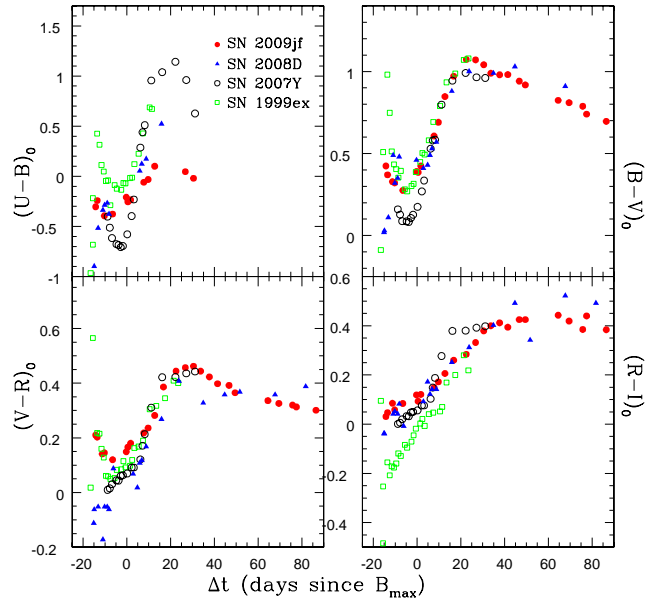}}
\caption[]{$(U-B)$, $(B-V)$, $(V-R)$ and $(R-I)$ colour curves of SN 2009jf 
compared with those of
SN 2008D, SN 2007Y and SN 1999ex.}
\label{fig_col}
\end{figure}

The colour evolution of SN 2009jf is plotted in Figure \ref{fig_col}. The 
colour curves of SN 2008D, SN 2007Y and SN 1999ex are also included in the 
figure for comparison. The colour curves have been corrected for total 
reddening values of  $E(B-V)$ of 0.112 for SN 2009jf, 0.65 for SN 2008D 
(Mazzali  et al. 2008), 0.112 for SN 2007Y \citep{strit09} and 0.3 
for SN 1999ex \citep{strit02}.  The reported magnitudes of SN 2007Y 
are in the $u^{'}$, $g^{'}$, $B$, $V$, $r^{'}$ and  $i^{'}$ bands. These 
magnitudes have been transformed to the $UBVRI$ system using transformation 
equations given in \cite{jester05}. The $(U-B)$, $(B-V)$ and $(V-R)$ 
colour curves of SN 2009jf evolve from red to blue in the pre-maximum epoch. 
This colour change can be attributed to an increase in the photospheric 
temperature with brightening of the supernova in the pre-maximum phase. The 
$(B-V)$ colour attains a value of 0.25 mag at $\sim$ 5 days before maximum in 
$B$ band, after that it monotonically becomes redder till $\sim$ 20 days after  
$B$ maximum, indicating cooling due to envelope expansion. It again starts 
becoming blue at later epochs. The $(V-R)$ colour follows a similar trend, 
while the $(R-I)$ colour evolves towards red monotoically. The $(B-V)$ and 
$(V-R)$ colour evolution of  SN 2009jf is quite similar to that of SN 1999ex, 
while the $(U-B)$ colour is always bluer and the $(R-I)$ colour redder than 
SN 1999ex. While SN 2009jf is redder than both SN 2007Y and SN 2008D in the 
pre-maximum epoch, the post maximum colour evolution is similar in all three 
SNe, except for the $(U-B)$ colour, which remains bluer in SN 2009jf. 

\section{Distance and Reddening} 

SN 2009jf is located at 54~arcsec west and 37~arcsec north of the nucleus of 
NGC 7479, at the edge of the outer arm of the host galaxy. From the infrared 
dust maps of \cite{schlegel98}, the Galactic interstellar reddening in the 
direction of NGC 7479 is E$(B-V)_{\rm{Gal}}$ = 0.112 mag. The spectrum of 
SN 2009jf obtained close to maximum light shows the presence of weak Na\,ID 
absorption from the Milky Way. We do not detect any Na\,ID absorption due to 
the host galaxy. The low reddening of the supernova is also evident from its 
optical colours (see Figure \ref{fig_col}). We therefore conclude there is no 
additional extinction due to the host galaxy and use a value of $E_(B-V)=0.112$
mag for extinction correction. 

The radial velocity of NGC 7479, corrected for Local Group infall onto the 
Virgo Cluster is 2443 km s$^{-1}$ (LEDA), which implies a distance modulus of 
$32.70\pm 0.18$ for an $H_{0}$ value of 71$\pm$6 km sec$^{-1}$ Mpc$^{-1}$. This 
leads to a distance of $34.66\pm 2.9$~Mpc for NGC 7479. The errors in distance 
modulus and distance are estimated taking into account the uncertainty in 
$H_{0}$. The redshift independent distance estimate using Tully-Fisher relation 
is $33.85\pm 3.1$ (NED), which is in close agreement with the distance 
estimates using the radial velocity. We use the mean of the two estimates, 
$34.25\pm 4.2$ Mpc as the distance to NGC 7479 for further analysis.
 
\section{Absolute magnitude, bolometric light curve and mass of $^{56}$Ni}

The absolute peak magnitudes estimated using a distance of 34.25 Mpc and a 
reddening correction for an $E(B-V)$ of 0.112 mag are listed in Table 
\ref{tab_peak}. The errors in the absolute magnitudes have been estimated using 
uncertainties in the peak magnitude and the distance modulus of the host galaxy.
Comparing the absolute magnitude of SN 2009jf with the absolute magnitude 
distribution of other SNe Ib (\cite{richardson06} and references therein),  
SN2009jf lies close to the mean of the  distribution. It is fainter than the 
extremely luminous type Ib supernova SN 1991D \citep{maza89}, and comparable 
in brightness to SN 1984L, SN1990I \citep{elmhamdi04}, SN1999ex 
(\citealt{strit02}, \citealt{hamuy02}) and SN 2000H \citep{krisciunas00}. 
SN 2009jf is $\sim$ 1.5 magnitude brighter than SN 2007Y \citep{strit09} 
and $\sim 1$ magnitude brighter than SN 2008D, which was associated with the 
X-ray transient 080909 \citep{modjaz09}. 

\begin{figure}
\resizebox{\hsize}{!}{\includegraphics{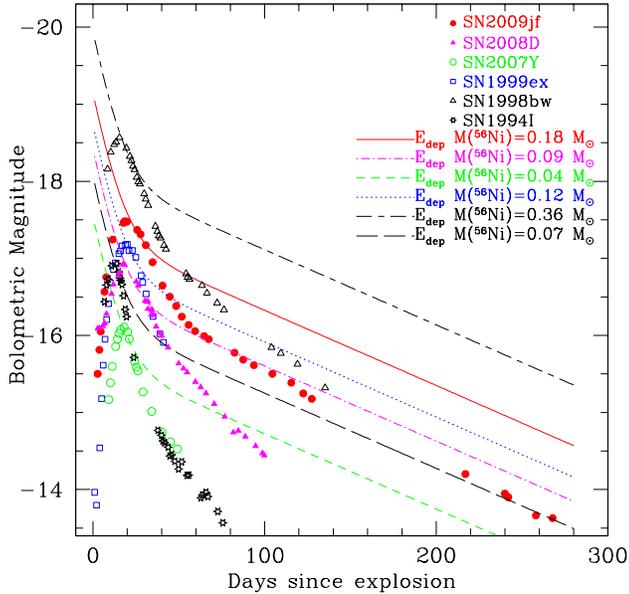}}
\caption[]{Bolometric light curve of SN 2009jf. Also plotted in the figure, for
comparison, are the bolometric light curves of SN 2008D, SN 1999ex, SN 1998bw, SN 2007Y 
and SN 1994I. The continuous curves correspond to the rate of energy production
for different masses of $^{56}$Ni sysntesized during the explosion, based on
the analytical formulation by \cite{nadyozhin94}}.
\label{fig_bol}
\end{figure}

The quasi-bolometric light curve of SN 2009jf is estimated using the reddening
corrected $UBVRI$ magnitudes presented here. The reddening corrected magnitudes 
were converted to monochromatic fluxes using the zero points from \cite{bessel98}.
 The quasi-bolometric fluxes were derived by fitting a spline curve to the $U$, 
$B$, $V$, $R$ and $I$ fluxes and integrating over the wavelength range 3100~\AA\
to 1.06$\mu$m, determined by the response of the filters used. There are a few 
missing magnitudes in the $U$ band light curve, which were estimated by 
interpolating between the neighbouring points. The quasi-bolometric light curve 
of SN 2009jf plotted in Figure \ref{fig_bol} is compared with the bolometric 
light curves of type Ib supernovae SN 2008D, SN 2007Y, SN 1999ex,  type 
Ic supernova SN 1994I and broad lined type Ic SN 1998bw,  also plotted in the 
figure. The bolometric light curves 
of supernovae SN 2008D, SN 2007Y, SN 1999ex and SN 1994I were constructed in a 
manner similar to SN 2009jf. The bolometric light curve of SN 2007Y \citep{strit09}, 
SN 1999ex \citep{strit02} 
and SN 1994I \citep{richmond96} are based on the published $UBVRI$ magnitudes, 
while the bolometric light curve of SN 2008D includes the {\it Swift} UVOT 
(U-band) and NIR data also \citep{tanaka09}. The bolometric light curve of SN 1998bw is
taken from \cite{patat01} which includes optical and NIR data. A total reddening $E(B-V)$ 
of 0.45, 0.3, 0.11, 0.65, 0.06 mag and, distances of 8.32, 48.31, 19.31 31.0 
and 37.8 Mpc were 
adopted for SN 1994I, SN 1999ex, SN 2007Y, SN 2008D and SN 1998bw, respectively.

The bolometric light curve of SN 2009jf is fainter than SN 1998bw and 
brighter than all other type Ib/c 
supernovae in comparison. Adding a conservative uncertainty of $\pm0.2$, mainly 
due to the uncertainty in $H_{0}$, the peak bolometric magnitude for SN 2009jf 
is estimated as $-17.48\pm0.2$ mag, which is $\sim 1.4$ magnitude brighter than 
SN 2007Y and $\sim 0.4$ magnitude brighter than SN 1999ex. The contribution of 
NIR and UV fluxes to the bolometric flux for type Ib/c supernovae is not well 
constrained. For SN 2007Y \cite{strit09} have estimated that close 
to the peak brightness, $\sim 70\%$ of the flux is in the optical bands, 
$\sim 25\%$ in the UV bands and $\sim 5\%$ in the NIR bands. However,
by two weeks past maximum the UV contribution comes down to $< 10\%$ and NIR 
contribution rises up to $\sim 20\%$. In the case of SN 2008D, the bolometric 
flux has an NIR contribution of about $< 24\%$ at $\sim 12$ days after 
maximum light in $V$ \citep{modjaz09}. Thus, the UV and NIR bands, together, 
contribute as much as $\sim30\%$ to the bolometric flux. Even without 
considering the UV and NIR contribution to the bolometric light curve, it is 
seen that SN 2009jf is 0.6 magnitude brighter than SN 2008D at peak.
Further, as seen in the $UBVRI$ light curves, the decline rate of bolometric 
light curve of SN 2009jf is slower than other supernovae in comparison. While 
the initial decline rate of SN 2009jf is comparable to SN 2008D, it is much 
slowler than SN 2008D in the later phases. The slope of the bolometric light 
curve $\sim$ 45 days after $B$ maximum is found to be 0.013 mag day$^{-1}$. 
For SN 2008D the same quantity is 0.023 mag day$^{-1}$. 

 The mass of $^{56}$Ni synthesized during explosion can be estimated following 
the principle that the bolometric luminosity,  L$_{bol}$ at maximum light is proportional to the
instantaneous rate of radioactive  decay \citep{arnett82}. The simplified formulation of 
Arnett's rule, to estimate mass of $^{56}$Ni,  $M_{Ni} = L_{bol}/ \alpha${\it\.S}  as 
proposed by \cite{nugent95}  involves $\alpha$, the ratio of bolometric to 
radioactivity luminosity  and \.S, the radioactivity luminosity per unit 
nickel mass, which depends on  the rise time of the supernova to maximum light.     
 The  peak $UBVRI$ bolometric luminosity  for SN 2009jf is estimated as 
${3}^{+0.6}_{-0.5}$$\times$10$^{42}$ erg sec$^{-1}$, the quoted uncertainty 
is mainly due to the uncertainty in $H_{0}$. 
SN 2009jf was discovered 
around 17.5 days before maximum in $B$ band, indicating the mimimum rise time
is around 18 days. Assuming a rise time of $19\pm 1$ days, the mass of 
$^{56}$Ni is estimated to be ${0.16}^{+0.03}_{-0.03}$ M$_\odot$ for SN 2009jf. 
It is worth mentioning here that $\alpha$,  the ratio of bolometric to radioactivity 
luminosity, which 
takes into account the possible radiation transport effects is assumed to be
unity. 

Another way for estimating mass of $^{56}$Ni synthesized during the explosion 
is to fit the energy deposition rate via \ $^{56}$Ni $\rightarrow$ $^{56}$Co chain, to 
the observed bolometric light curve. The total rate of energy production via 
\ $^{56}$Ni $\rightarrow$ $^{56}$Co chain estimated using the analytical formula by 
\cite{nadyozhin94}, for different values of mass of $^{56}$Ni \ synthesized during
the explosion is plotted with the bolometric light curves in Figure 
\ref{fig_bol}. The  energy deposition rate corresponding to $^{56}$Ni \ mass of 
0.18 M$_\odot$ fits the initial decline of the quasi-bolometric light curve 
of SN 2009jf.   The mass of $^{56}$Ni estimated using Arnett's law and the 
energy deposition rate are in good agreement with each other, and in what follows,
an average of the two estimates, 0.17 M$_\odot$ is taken as the mass of $^{56}$Ni synthesized in the
explosion.  In the above estimates of $^{56}$Ni mass, the contribution from UV 
and NIR bands to the bolometric luminosity have not been included. As discussed 
earlier, the contribution from UV and NIR bands to the bolometric luminosity 
at any given time is about $30\%$. Including this contribution to the 
bolometric luminosity, the estimated
mass of $^{56}$Ni synthesized in the explosion will increase by $\sim30\%$.

\section{Spectral evolution} 

Our spectroscopic observations began 15 days before $B$ maximum and continued 
till 99 days after $B$ maximum, when the object went in solar conjunction. 
Subsequently, two spectra were obtained in the nebular phase, at 229 days and 
251 days after $B$ maximum.

\subsection{Early phase}
\begin{figure}
\resizebox{\hsize}{!}{\includegraphics{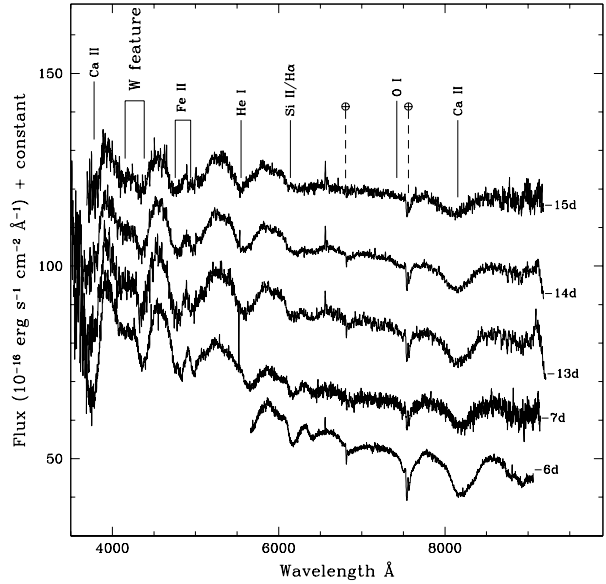}}
\caption[]{Pre-maximum spectral evolution of SN 2009jf during -15 to -6 days with respect to 
maximum in $B$ band.}
\label{fig_early_spec}
\end{figure}

\begin{figure}
\resizebox{\hsize}{!}{\includegraphics{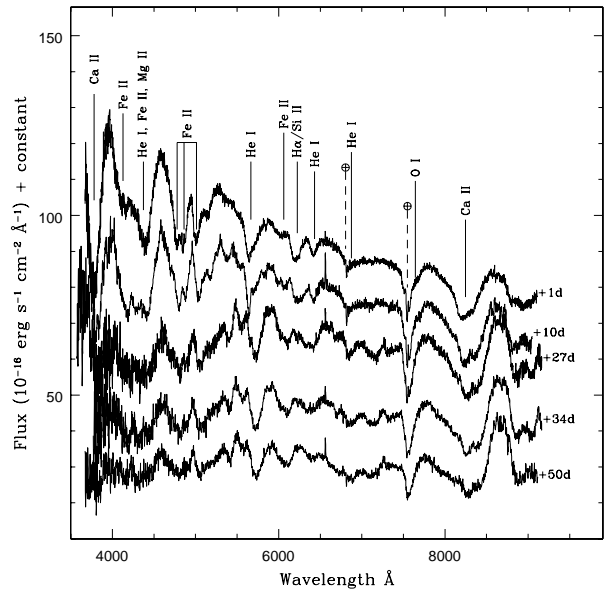}}
\caption[]{Spectral evolution of SN 2009jf around maximum and immediate post-maximum phase.}
\label{fig_max_spec}
\end{figure}

\begin{figure}
\resizebox{\hsize}{!}{\includegraphics{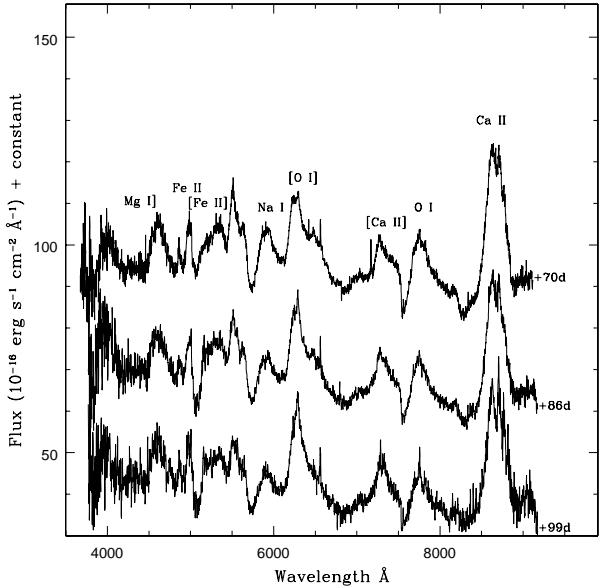}}
\caption[]{Spectral evolution of SN 2009jf during +70 to +99 days with respect to 
maximum in $B$ band.}
\label{fig_postmax_spec}
\end{figure}

The spectral evolution of SN 2009jf is plotted in Figures \ref{fig_early_spec},
\ref{fig_max_spec} and \ref{fig_postmax_spec} . Our 
first spectrum is one of the earliest spectrum for type Ib supernovae, along 
with SN 2007Y \citep{strit09} and SN 2008D \citep{modjaz09}. All spectra have 
been corrected for the heliocentric velocity 2381 km sec$^{-1}$ of the host galaxy. 

The pre-maximum spectra of SN 2009jf are plotted in   Figure
\ref{fig_early_spec}. These spectra are characterized by a broad P-Cygni 
profile, indicative of high expansion velocity of the ejecta. The first 
spectrum taken on JD 2455104.16 (15 days before $B$ maximum) shows 
distinctive broad absorption due to He\,I 5876~\AA\ with an expansion 
velocity of $\sim 16\,300$ km sec$^{-1}$ and possible contribution from Na\,ID 
5890, 5896~\AA. The other well developed features seen in the first spectrum 
are due to Ca\,II H \& K  3934, 3968~\AA, Fe\,II between 4100 to 5000~\AA, 
Si\,II/H$\alpha$ at $\sim$ 6250~\AA, O\,I 7774~\AA\ and Ca\,II NIR triplet between 
8000 and 9000~\AA. Clear signature of other  He\,I lines 4471~\AA\ (possibly 
blended with Fe\,II 4924~\AA\ and Mg\,II 4481~\AA), 5015, 6678 and 7065 is 
present in the spectrum taken $\sim$ 13 days before $B$ maximum. The He\,I 
7065~\AA\ line is affected by the telluric H$_{2}$O band. The first spectrum 
also shows the double absorption, the ``W'' feature, at $\sim 4000$~\AA\ seen 
in the very early spectra of the type II supernova SN 2005ap \citep{quimby07}, 
the type Ib supernova SN 2008D \citep{modjaz09} and the type IIb supernova 
SN 2001ig \citep{silverman09}.  This feature is identified with Fe complexes
\citep{mazzali08}, or as a combination of C\,III, N\,III and O\,III lines at 
high velocities (\citealt {modjaz09},  \citealt{silverman09}). \cite{tanaka09} 
have investigated the presence of C\,III, N\,III and O\,III lines in the early 
spectrum of SN 2008D using a Monte Carlo spectrum synthesis code, and do not 
find a large contribution from these ionization states. They conclude that 
ionization by the photospheric radiation only is not enough for the observed 
features to be due to these doubly ionized lines.  

The continuum becomes bluer as the supernova evolves towards maximum, as also 
indicated by the colour curves (Figure \ref{fig_col}).

\begin{figure}
\resizebox{\hsize}{!}{\includegraphics{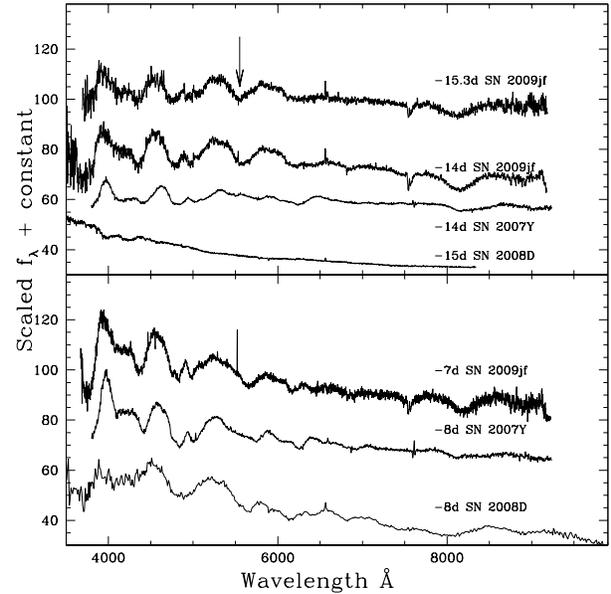}}
\caption[]{Comparison of pre-maximum spectra of SN 2009jf with SN 2008D and  SN 2007Y.
 Note the early emergence of He\,I line in SN 2009jf, marked by arrow.}
\label{fig_spec_comp1}
\end{figure}

The pre-maximum spectra of SN 2009jf are compared with those of SN 2007Y and 
SN 2008D at similar epochs and shown in Figure \ref{fig_spec_comp1}. The 
spectrum of SN 2009jf at 15 days before $B$ maximum (top panel), shows
well developed absorptions at 4100--5500~\AA\ due to Fe\,II and He\,I 5876~\AA. 
In comparison, SN 2008D shows a nearly featureless spectrum around the same 
time. While the Fe\,II features are clearly identifiable in SN 2007Y, the He\,I
5876~\AA\ feature is absent. The spectra of SN 2009jf and SN 2007Y at $\sim$ 7 
days before $B$ maximum (lower panel) are very similar, except for the 
differences in the expansion velocities, while in SN 2008D, lines due to He\,I 
are identified, but Fe\,II lines are still not well developed.  

The post-maximum spectra are shown in Figures
\ref{fig_max_spec} and \ref{fig_postmax_spec}. The spectrum of SN 2009jf obtained 
1 day after $B$  maximum shows a bluer continuum, with 
well developed lines due to Ca\,II H \& K, He\,I, Fe\,II and a broad P-Cygni profile 
of Ca\,II near-IR triplet. The prominent absorption at $\sim$ 6250~\AA\ in the 
pre-maximum phase weakens, and is not seen in the spectra beyond day $+$10. The 
continuum of the post-maximum spectrum on day $+$27 again becomes redder. Later on,
the evolution of the spectrum is slow, with further suppression of the flux in 
blue and an increase in the flux of Ca\,II NIR triplet. 

\begin{figure}
\resizebox{\hsize}{!}{\includegraphics{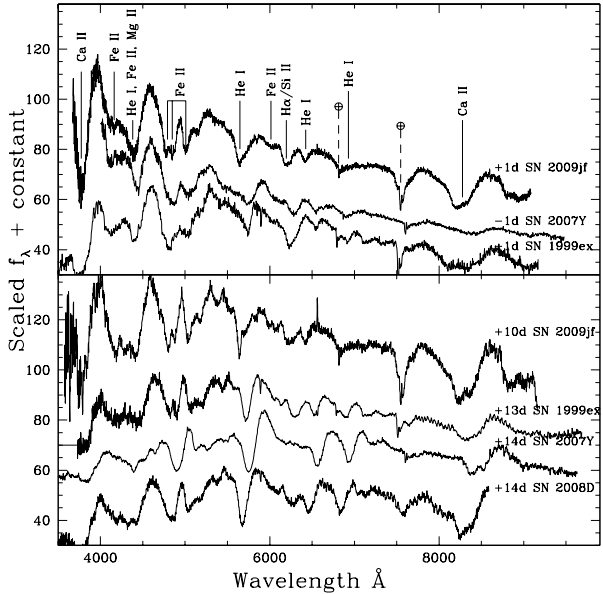}}
\caption[]{Spectral comparison of SN 2009jf around maximum and immediate
post-maximum phase.}
\label{fig_spec_comp2}
\end{figure}

\begin{figure}
\resizebox{\hsize}{!}{\includegraphics{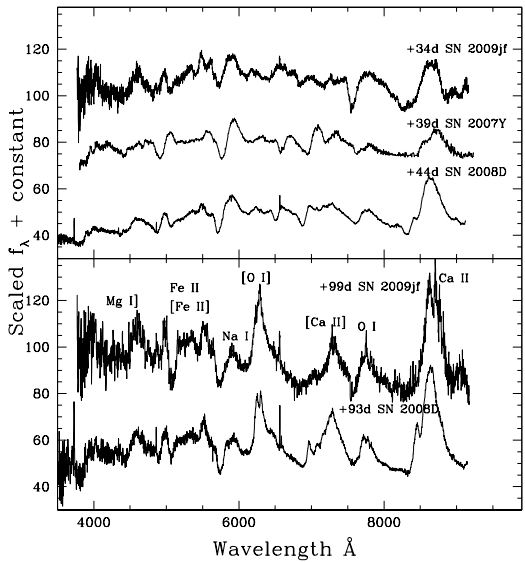}}
\caption[]{Spectral comparison of SN 2009jf at later phases.}
\label{fig_spec_comp3}
\end{figure}

The spectra of SN 2009jf, SN 2007Y and SN 1999ex close to maximum are plotted
in Figure \ref{fig_spec_comp2} (top panel). The main features in the spectra 
are identified and marked in the figure. While the general characteristics of 
the spectrum in the three SNe are similar, it is seen that the Fe\,II lines 
around $~\sim 5000$~\AA\ are somewhat underdeveloped in SN 1999ex. In the
phase $\sim 10$~days after $B$ maximum (Figure \ref{fig_spec_comp2}: lower 
panel), the spectral features in SN 2009jf appear to be narrower compared to 
the other supernovae. Around a month after maximum, all three supernovae show
identical features (Figure \ref{fig_spec_comp3}: top panel).

Forbidden emission lines of [O\,I] 6300-6364~\AA\ and [Ca\,II] 7291, 7324~\AA\ 
are seen in the spectrum of day $+70$ (Figure \ref{fig_postmax_spec}), 
marking the onset of the nebular phase. The spectra of $+86$ and $+99$ days 
after $B$ maximum show a strengthening of the [O\,I] and [Ca\,II] features.
The other features identified in these spectra are  Mg\,I] 4570~\AA, [Fe\,II] blend
at 5200~\AA, Na\,I doublet 5890, 5896~\AA, O\,I 7774~\AA\ and the blend at 
$\sim 8700$~\AA, which has contributions from O\,I 8446~\AA, Ca\,II 8498--8662~\AA,
and [C\,I] 8727~\AA\ \citep{fransson89}. The features have been identified in 
Figure \ref{fig_spec_comp3} (bottom panel). The [O\,I] profile at this phase is
single peaked and asymmetric, with the emission peaking redwards. The asymmetry
is more evident in the spectrum of day $+99$. In comparison, the $+93$ day spectrum 
of SN 2008D shows a symmetric double peaked [O\,I] line.   

\subsection{Nebular phase}

\begin{figure}
\resizebox{\hsize}{!}{\includegraphics{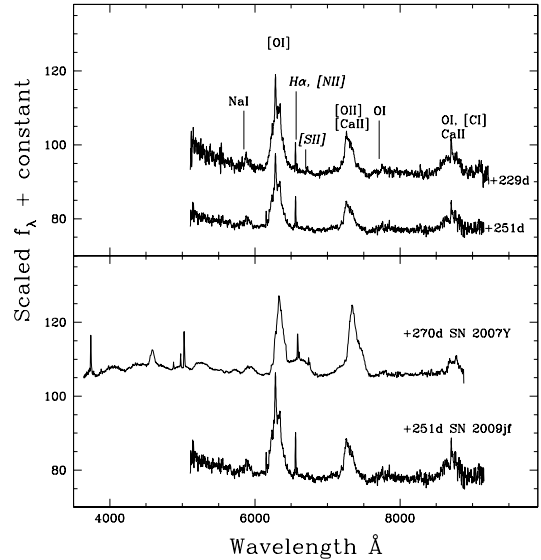}}
\caption[]{Nebular spectra of SN 2009jf (top). H$\alpha$+[N\,II] and [S\,II] features 
(marked in {\it italics} ) from the underlying H\,II region are also identified in the spectra.
 The bottom panel shows
the comparison with the nebular spectrum of SN 2007Y.}
\label{fig_neb}
\end{figure}

The spectra taken  $229$ and $251$ days after maximum are presented 
in Figure \ref{fig_neb}. The spectrum during these days is dominated by the
[O\,I] and [Ca\,II] features. The [Ca\,II] emission is blended with [Fe\,II] lines at
7155, 7172, 7388 and 7452~\AA\ and possibly with [O\,II] 7320, 7330~\AA\ 
\citep{strit09}. Na\,I 5890, 5896~\AA\ doublet, O\,I 7774~\AA, and the 8700~\AA\ 
blend are also identified in these spectra, but with a decreased strength 
compared to the spectra of 86 and 99 days after maximum. The FWHM of a Gaussian 
fit to the [O\,I] emission line indicates a velocity dispersion of $\sim 7300$ km sec$^{-1}$, 
whereas the corresponding velocity dispersion estimated for the [Ca\,II]
line is $\sim 6500$ km sec$^{-1}$. The FWHM velocities measured for SN 2009jf are 
higher than those  for SN 2007Y \citep{strit09}, SN 1996N \citep{sollerman98},  
SN 1985F \citep{schlegel89} and the sample of type Ib supernovae discussed in
\citep{matheson01}, at similar epochs. The blend at $\sim 8700$~\AA\ has a FWHM 
of $\sim 8000$ km sec$^{-1}$ on day $+229$, which decreases to $\sim 7300$ km 
sec$^{-1}$ on day $+251$. The Ca\,II NIR/[Ca\,II] line ratio measured in the two nebular 
spectra indicates that the Ca\,II NIR is getting weaker as compared to the [Ca\,II] 
line, a feature noticed in SN 1996N \citep{sollerman98} and interpreted as due 
to decreasing density \citep{filipenko90}. However, the blend at 8700~\AA \ has
contribution from O\,I, Ca\,II and [C\,I], hence the measured velocity and flux using 
this blend must be viewed with caution. 

The emission line profiles are multi-peaked and asymmetric. The [O\,I] line shows 
a sharp and stronger blue peak. A similar profile is clearly apparent in the 
[Ca\,II] line of day $+229$. The sharp emission component appears to be present 
in all the emission lines. In addition to the broad features due to the 
supernova ejecta, the nebular spectra also show narrow lines due to H$\alpha$, 
[N\,II] 6548, 6583~\AA\ and [S\,II] 6717, 6731~\AA, originating from the underlying 
H\,II region.

\subsection{Expansion velocity of the ejecta}

\begin{figure}
\resizebox{\hsize}{!}{\includegraphics{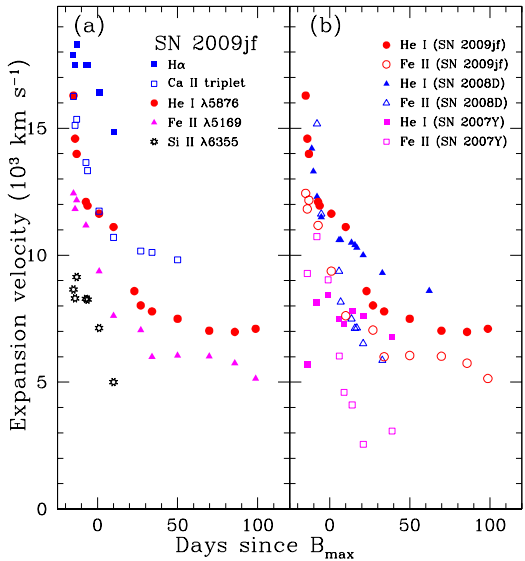}}
\caption[]{Left: Temporal velocity evolution of the prominent ions in SN 2009jf 
spectra. Right: Comparison of the He\,I and Fe\,II lines velocities in SN 2009jf
with those observed in SN 2007Y and SN 2008D.}
\label{fig_vel}
\end{figure}

Expansion velocities of the prominent features seen in the spectra are estimated
by fitting a Gaussian profile to the minimum of the absorption trough in the 
redshift corrected spectra. The velocity evolution of the prominent ions, seen 
in the spectra of SN 2009jf, is plotted in Figure \ref{fig_vel}a. The expansion 
velocity of He\,I 5876~\AA\ line, determined using the pre-maximum spectra, 
rapidly decreases from a value of $\sim 16000$ km sec$^{-1}$ on day $-$15 to 
$\sim 12000$ km sec$^{-1}$ close to $B$ maximum. The velocity further declines in 
the post-maximum phase and levels off at $\sim 7000$ km sec$^{-1}$. The expansion 
velocity of Fe\,II 5169~\AA\ feature remains low as compared to that of He\,I 5876~\AA\ 
line all through its evolution, and also declines at a slower rate, as seen in 
most type Ib supernovae \citep{branch02}. The expansion velocities of Ca\,II 
near-IR triplet follows the evolution of He\,I 5876~\AA\ line, with a marginally
higher velocity in the pre-maximum phase. While the velocity of the He\,I line
decreases to $\sim 7000$ km sec$^{-1}$ in the pre-maximum phase, the velocity of 
Ca II near-IR triplet remains higher at $\sim 10000$ km sec$^{-1}$. 

 The feature seen at 6250~\AA\ in the spectrum of some type Ib supernovae, has 
been identified with  H$\alpha$ in SN 1954A, SN 1999di, SN 2000H \citep {branch02}, 
SN 2005bf \citep{anupama05},  or with Si\,II in SN 1999ex \citep{hamuy02}, or as a blend of
C\,II 6580~\AA\ and H$\alpha$ in SN 1999dn \citep{deng00}, or as a blend of Si\,II 
and H$\alpha$ in SN 2008D \citep{tanaka09}. In the case of SN 2005bf, 
identification of the
6250~\AA\ feature with H$\alpha$ was supported by the presence of a blueshifted 
H$\beta$ line in the early spectrum \citep{anupama05}. Similarly, in SN 2007Y 
\cite{strit09} have identified the 6250~\AA \ feature with H$\alpha$ possibly 
blended with Si\,II based on the presence of H$\alpha$ in the nebular spectrum 
and a possible presence of H$\beta$ in the early spectrum.  The presence of a high 
velocity  H$\alpha$ feature in the early phase is usually accompanied by the presence of 
a broad shoulder redward of the [O\,I] 6300-6364~\AA\ feature, as seen in the 
spectrum of SN 2007Y.  The presence of H$\alpha$
absorption at early times and strong H$\alpha$ emission in the late phase
led \cite{maurer10} to reclassify SN 2007Y as type IIb. Identifying the
feature seen at 6250~\AA\ in the spectra of SN 2009jf with H$\alpha$ indicates 
expansion velocities much higher than  He\,I, while identification with 
Si\,II indicates velocities that are lower than Fe\,II (see Figure \ref{fig_vel}).  
Further, in the nebular spectra of SN 2009jf presented here, the broad
shoulder redward of [O\,I] 6300-6364~\AA\ feature is also not seen. We therefore 
prefer to identify the  6250~\AA\ feature with Si\,II.

The He\,I 5876~\AA\ and Fe\,II 5169~\AA\ line velocities estimated for SN 2009jf
are compared with those of SN 2008D and SN 2007Y in Figure \ref{fig_vel}b. The 
velocity evolution of SN 2009jf is similar to that of SN 2008D. The He\,I 
velocity in both these objects is higher than the Fe\,II line velocity, whereas
SN 2007Y has a higher Fe\,II velocity \citep{strit09}. 
 The He\,I velocity in SN 2009jf during the pre-maximum and early 
post-maximum phases is higher than SN 2008D, but $\sim$ 20 days after $B$ 
maximum, the velocity in SN 2009jf becomes lower than SN 2008D. On the other
hand, the Fe\,II line velocity is initially lower in SN 2009jf and becomes
similar to SN 2008D at later epochs. SN 2007Y has lower velocities compared to
both SN 2009jf and SN 2008D at all phases.
    
\section{The Oxygen mass}
The nebular spectra of type Ib supernovae are dominated by [O\,I] emission,
considered the prime cooling path during the late phases (\citealt{uomoto86}, 
\citealt{fransson87}). The absolute flux of this line  can be used to estimate the 
mass of neutral oxygen producing the line emission, following the expression by 
\cite{uomoto86}, 

\begin{equation}
M_{O} = 10^{8}\times D^{2}\times {\rm{F([O\,I])}} \times \exp {(2.28/T_4)} \,\,\, ,
\end{equation}
where $M_{\rm{O}}$ is the mass of neutral oxygen in M$_\odot$, $D$ is the distance 
to the supernova in Mpc, F([O\,I]) is the flux of the [O\,I] line in ergs sec$^{-1}$
and $T_{4}$ is the temperature of the oxygen emitting region in units of $10^{4}$ K.
  The above equation holds in the high density regime ($N_{e} \geq 10^{6}$ cm$^{-3}$),
which is met in the ejecta of type Ib supernovae (\citealt{schlegel89}, 
\citealt{elmhamdi04}, \citealt{gomez94}). An estimate of the temperature of the
line emitting region can be made using the [O\,I] 5577/6300-6364 flux ratio. 
[O\,I] 5577~\AA\ line is not detected in the nebular spectrum of SN 2009jf. This
implies a limit on the [O\,I] 5577/6300-6364 flux ratio of $\leq 0.1$. At this
limit, the emitting region should either be at a relatively low temperature 
(T$_{4} \leq 0.4$) for the high density limit, and/or at low electron density
(n$_{e} \leq 5\times 10^{6}$ cm$^{-3}$) if T$_{4}$ = 1 \citep{maeda07}.
 
\cite{elmhamdi04} estimate a temperature of $\sim 3200-3500$~K for SN 1990I at 
$\sim$ 237 days after maximum light, using an upper limit of the flux of 
[O\,I]$\lambda$5577 line, while \cite{schlegel89}, have estimated the mass of 
oxygen in type Ib supernovae SN 1984L and SN 1985F by assuming T$_{4}$ = 0.4.
In all cases, a density $\ga 10^6$ cm$^{-3}$ is assumed. Assuming that the
high density regime is valid for SN 2009jf also, a value of T$_4=0.4$ appears
to be a good approximation.
Using the [O\,I] flux of 3.74$\times$10$^{-14}$ erg sec$^{-1}$ cm$^{-2}$ measured in the 
spectrum of day $+251$, and the assumed distance of 34.25 Mpc, the mass of oxygen
is estimated to be 1.34 M$_\odot$.   A weak line at $\sim$ 7750~\AA\ 
is present in the nebular spectra of SN 2009jf and is identified with O\,I 
7774~\AA\ line, following \cite{mazzali10}. The presence of O\,I 7774~\AA\ line
in the spectrum is indicative of the presence of ionized oxygen also, as this 
line is mainly due to recombination \citep{begelman86}. \cite{mazzali10} have 
shown that the mass of oxygen required to produce  [O\,I] 6300-6364~\AA\ and
O\,I 7774~\AA\  lines together is higher than that is required to produce only 
[O\,I] 6300-6364~\AA\ line. Thus, the oxygen mass
estimate of 1.34 M$_\odot$ may be considered as a lower limit of the 
total mass of oxygen ejected during the explosion.

\section{Discussion} 

The light curve and spectral evolution of SN 2009jf show some peculiarities
compared to other SNe Ib. The light curves indicate a post-maximum decline that
is slower compared to other type Ib supernovae. This slow decline continues even
during the late phases, making the light curve of SN 2009jf broad.  
The absolute $V$ magnitude at peak is comparable to the mean of the absolute 
magnitude distribution of type Ib supernovae.  Using the bolometric light curve 
and the energy deposition rate via $^{56}$Ni $\rightarrow$ $^{56}$Co, the mass 
of $^{56}$Ni synthesized during the explosion is estimated to be 0.17 M$_\odot$. 
SN 2009jf shows a very early emergence of He\,I lines in the spectrum. He\,I 5876~\AA\
line is identified in the first spectrum obtained 15.3 days before $B$ maximum. 
Other lines due to He\,I at 4471~\AA, and 6678~\AA \ were identified in the $-13$ 
day spectrum. Further, the expansion velocity estimated using He\,I line 
$\sim 16,000$ km sec$^{-1}$, indicating that helium is excited at high velocity. 
In case of SN 2008D, He\,I lines became apparent around 11.5 days before $B$
maximum, and were prominent only around 5 days before $B$ maximum \citep{modjaz09}. 
The He\,I lines seen in the spectra of type Ib supernovae require non-thermal 
excitation and ionization, as the temperature present in the ejecta is too low 
to cause any significant absorption \citep{lucy91}. $\gamma$-rays, emitted by 
newly synthesized $^{56}$Ni during the explosion, accelerate electrons that 
act as a source of non-thermal excitation for He (\citealt{harkenss87}, \citealt{lucy91}). 
For exciting helium at such a high velocity as seen in SN 2009jf, the 
$\gamma$-rays need to be close to the helium layer, which can be possible 
either through the escape of $\gamma$-rays from the $^{56}$Ni dominated region,
or through some large scale instability causing  substantial mixing of 
$^{56}$Ni to the outer layers. The slower decline of the light curves of 
SN 2009jf gives an indication that it has a massive ejecta and the probability 
of $\gamma$-rays escaping will be low.  Though substantial mixing of different 
inner layers appears to be the most probable way for an early excitation of He 
at high velocities, the possibilty of some $\gamma$-rays reaching the He layer and 
exciting it cannot be ruled out, especially since SN 2009jf is a rather luminous
supernova.

The profile of [O\,I] 6300-6364~\AA\ feature in the nebular spectrum is multi-peaked
and asymmetric with a sharp, stronger blue peak. The peak of this feature is 
blueshifted by $\sim 30$~\AA\ around day +86, which reduces to a blueshift of
$\sim 15$~\AA\ by day +99. Such observed blueshifts are explained as a result
of residual opacity in the core of the ejecta \citep{taubenberger09}. The 
asymmetric and multi-peaked profile seen at phases later than 200 days can be 
produced by additional components of arbitrary width and shift with respect to 
the main component. Such profiles are indicative of an ejecta with large-scale 
clumping, a single massive blob, or a unipolar jet. The asymmetric [O\,I] line 
profile of SN 2009jf with a stronger blue peak is very similar to the line 
profiles of SNe 2000ew and 2004gt. \cite{taubenberger09} have explored a 
possible configuration which can give rise to this asymmetric line profile, and 
interperted the profile as originating from the deblended 6300~\AA\ and 
6364~\AA\ lines of a single narrow, blueshifted component.  \cite{maurer10} 
have shown that the profile of [O\,I] 6300-6364~\AA\ doublet is likely to be
influenced by H$\alpha$ absorption. If hydrogen concentration is located around
$\sim$ 12000 km sec$^{-1}$, it  causes a split of the [O\,I] 6300-6364~\AA\ 
doublet, leading to a double-peaked oxygen profile. Neither scenarios account
for the stronger blue peak of the 6300~\AA\ line. \cite{taubenberger09} explain
the stronger blue peak  with a complex ejecta structure with additional 
blueshifted emission on top of an otherwise symmetric profile. Alternatively, 
the asymmetry in the 
profile is explained by a damping of the redshifted emission component in an 
originally toroidal distribution, caused by the optically thick inner ejecta. 
The light curve evolution of SN 2009jf indicates the presence of an ejecta more 
massive than other stripped core collapse supernovae. Hence, it is quite likely 
that in the case of SN 2009jf also the redward component is damped by an 
optically thick inner ejecta.  It should however be noted that asymmetric 
and multi-component profiles cannot be reproduced within spherical symmetry 
\citep{mazzali05, maeda07}. This needs further investigation with observations 
at phases later than presented here, as well as spectrophotometric observations
and detailed modelling. 

The brightness and width of Type Ib light curves are determined by the interplay
of nickel mass, opacity and $\gamma$-ray deposition. In general, a greater amount 
of $^{56}$Ni will make the light curve brighter. A more massive ejecta will have a larger
optical depth, and it will take longer for the trapped decay energy to diffuse 
through the envelope, which will broaden the light curve \citep{ensman88}. The 
time taken for the bolometric light curve to decline from peak to the moment 
when the luminosity is equal to $1/e$ times the peak luminosity (which is 
equivalent to a decline of 1.1 mag from peak), is known as the effective diffusion 
time $\tau_{\rm m}$. The effective diffusion time is related to the mass of the 
ejecta M$_{\rm ej}$ and the kinetic energy E$_{\rm k}$ of the ejecta  
$\tau_{\rm m} \propto \kappa_{\rm opt}^{1/2} M_{\rm ej}^{3/4} K_{\rm E}^{-1/4}$ 
\citep{arnett82}, where  $\kappa_{\rm opt}$ is optical opacity. The expansion 
velocity $v$ can be expressed in terms of M$_{\rm ej}$ and E$_{\rm k}$ 
as $v \propto M_{\rm ej}^{-1/2} K_{\rm E}^{1/2}$. The broad peak and slower 
decline rates of the light curves of SN 2009jf in comparison to other supernovae
indicate that SN 2009jf has a massive ejecta. Further, the broader emission 
lines at late phase indicates a larger explosion energy.  

There are several core-collapse supernovae for which the progenitor mass has 
been constrained using hydrodynamical modelling. With this approach, 
\cite{nomoto03} and \cite{nomoto04} constructed
the $E_{\rm K} - M_{\rm MS}$ diagram and introduced a hypernova branch.  
Recent updates of this approach include SN 1998bw \citep{maeda06}, 
SN 2008D \citep{tanaka09}, and SN 2003bg \citep{mazzali09}.
For the well studied bright 
hypernova SN 1998bw ($M_V=-19.35$ \cite{galama98}), the main sequence
 mass of the progenitor is constrained by \cite{maeda06} as $\sim$ 40 M$_\odot$.
Though SN 2009jf has a brighter peak compared to SN 2008D, the fact that the 
light curves of SN 2009jf around maximum and the initial decline rate 
$\Delta m_{15}(V)$ are similar to those of SN 2008D can be used to estimate 
the mass of the ejecta M$_{ej}$ and kinetic energy E$_{\rm k}$ of the ejecta, 
using SN 2008D as 
the reference, assuming the optical opacity $\kappa_{\rm opt}$ to be the same. The 
effective diffusion time $\tau_{\rm m}$ for SN 2009jf  and SN 2008D are estimated
to be 30 days and 26 days, respectively. The photospheric expansion velocity
estimated using the Fe\,II 5169~\AA\ line at maximum is $\sim$ 10000 km 
sec$^{-1}$, similar for both the objects. For SN 2008D, the mass of the 
ejecta M$_{\rm ej}$, the kinetic energy E$_{\rm k}$ and the progenitor mass 
have been derived by \cite{mazzali08}, \cite{soderberg08} and \cite{tanaka09}. 
\cite{mazzali08} could
reproduce the  spectral evolution and light curve  with a spherically symmetric
explosion energy  E$_{\rm k} = 6.0\times 10^{51}$ erg and ejected mass 
M$_{\rm ej} \sim$ 7 M$_\odot$ with a progenitor of mass $\sim$ 30 M$_\odot$
 while \cite{soderberg08} have arrived at 
E$_{\rm k} = 2-4 \times 10^{51}$ erg and M$_{\rm ej}$ = $3-5$ M$_\odot$, by applying
rescaling arguments.   \cite{tanaka09} have calculated 
the hydrodynamics of explosion and explosive nucleosynthesis for SN 2008D with
varying mass for the He core of the progenitor and concluded that the progenitor
star of SN 2008D had a He core mass $6-8$ M$_\odot$ prior to explosion. 
This corresponds to a main sequence mass of $M_{\rm MS} = 20-25$ M$_\odot$. 
The explosion energy and mass of ejecta for SN 2008D were estimated to be 
E$_{\rm k}=6.0\pm 2.5\times10^{51}$ erg and M$_{\rm ej} = 5.3\pm 1.0$ M$_\odot$, 
respectively. Thus, for SN 2008D the mass of ejecta  M$_{\rm ej}$ and explosion 
energy E$_{\rm k}$ range between $3-7$ M$_\odot$ and $2-6 \times 10^{51}$ erg, respectively. 
Using the observed photospheric velocity and the estimated
diffusion time for SN 2009jf, and treating  SN 2008D as a reference, we estimate  
M$_{\rm ej} = 4-9$ M$_\odot$ and E$_{\rm k} = 3-8\times 10^{51}$ erg
for SN 2009jf.  This indicates that SN 2009jf was an energetic explosion of a 
star having a mass similar, or somewhat more massive than the progenitor of 
SN 2008D ($M_{\rm MS} \ga 20-25$ M$_\odot$). The physical parameters of SN 2009jf 
may also be compared with those of the type IIb supernova SN 2003bg, which had 
an absolute peak magnitude of $M_V=-17.5$ \citep{hamuy09} and an oxygen mass
estimate of 1.3 M$_\odot$. \cite{mazzali09}  have estimated the physical
parameters for SN 2003bg based on detailed light curve and spectral modelling.  
The best fit model gives an ejected mass of $\sim 4.8$ M$_\odot$, kinetic 
energy $\sim 5\times 10^{51}$ erg and mass of $^{56}$Ni $\sim 0.15-0.17$ M$_\odot$. 
The mass of the progenitor for SN 2003bg is estimated as $20-25$ M$_\odot$. 
Our qualitative analysis of light curve and spectra of SN 2009jf hints towards 
a higher kinetic energy and a slightly more massive ejecta than SN 2003bg, and 
in turn a progenitor with $M_{\rm MS} \ga 20-25$ M$_\odot$. 

The mass of oxygen $M$(O) in the ejecta of the core collapse SNe is
very sensitive to the main-sequence mass $M_{\rm MS}$ of the
progenitor.  For $M_{\rm MS}$ = 15, 18, 20, 25, 30, and 40 M$_\odot$,
$M$(O) = 0.16, 0.77, 1.05, 2.35, 3.22, and 7.33 M$_\odot$,
respectively \citep{nomoto06}.  These values are obtained for E$_{\rm
k}=1.0\times10^{51}$ erg and the metallicity $z=0.02$, but are not so
sensitive to E$_{\rm k}$ and $z$.  In fact, for ($M_{\rm
MS}$/M$_\odot$, E$_{\rm k}/10^{51}$ erg) = (20, 10), (25, 10), and
(30, 20), and (40, 30), $M$(O)/M$_\odot$ = 0.98, 2.18, 2.74, and 7.05,
respectively \citep{nomoto06}.  Therefore, the lower limit of the
oxygen mass $M$(O) $\ga$ 1.34 M$_\odot$ estimated from the nebular
spectra is quite consistent with
the progenitor mass of $M_{\rm MS} \ga 20-25$ M$_\odot$ estimated from
the light curve shape and the photospheric velocities.

The [Ca\,II] 7291-7324/[O\,I]6300-6364 line ratio is also a good
diagnostic of $M_{\rm MS}$, because the mass of the explosively
synthesized Ca in the ejecta, $M$(Ca), is not sensitive to $M_{\rm
MS}$.  For $M_{\rm MS}$/M$_\odot$ = 15, 18, 20, 25, 30, and 40,
$M$(Ca)/10$^{-2}M_\odot$ = 0.40, 0.45, 0.37, 0.66, 1.6, and 1.6,
respectively \citep{nomoto06}.  Also, for ($M_{\rm MS}$/M$_\odot$,
E$_{\rm k}/10^{51}$ erg) = (20, 10), (25, 10), and (30, 20), and (40,
30), $M$(Ca)/10$^{-2}M_\odot$ = 0.50, 0.57, 0.93, and 1.4,
respectively \citep{nomoto06}.  This is in contrast to $M$(O), which
sensitively increases with $M_{\rm MS}$.  Thus a smaller [Ca\,II]/[O\,I] ratio 
indicates a  massive core. 

 The [Ca\,II] 7291-7324/[O\,I]6300-6364 emission line ratio for 
SN 2009jf is estimated as 0.51 and 0.49 using the nebular spectrum observed 
on days $+229$ and $+251$, respectively. For SN 2007Y, SN 1996N, 
SN 1990I and SN 1998bw, this ratio was found to be 1.0, 0.9, 0.7 and 0.5, 
respectively, at similar epochs (\citealt{elmhamdi04}, \citealt{strit09} and 
references therein). 
\cite{fransson89}  have theoretically calculated the [Ca\,II]/[O\,I] line ratio 
for progenitor masses of 15 and 25 M$_{\odot}$. The observed [Ca\,II]/[O\,I] 
ratio for SN 2009jf is very close to the ratio expected for a star with 
$M_{\rm MS}$= 25 M$_\odot$, as indicated by Model 1b in \cite{fransson89}. 

The estimates of the mass of
$^{56}$Ni synthesized during the explosion, the kinetic energy of explosion and 
the main sequence mass of the progenitor star places SN 2009jf between the 
normal core-collapse supernovae and the hypernovae branch in the $E_K - M_{MS}$ 
diagram of \cite{tanaka09}, at the upper end of the normal core-collapse 
supernovae branch. It is however to be noted that the [O\,I] line profile
during the nebular phase indicates asymmetry of the explosion. This can have 
some effect in the kinetic energy estimate, as shown by \cite {maeda06} and 
\cite{tanaka07} for SN 1998bw. A detailed modelling is therefore required for a
better estimate of the various parameters.    
 
\cite{itagaki09} suggest the progenitor could have undergone luminous blue
variable type mass loss events, based on their detection of a dim object at 
the location of the supernova on three occasions.  Pre-supernova images of 
the host galaxy obtained in the ultraviolet by the {\it Swift} satellite, and
available in the {\it Swift} data archives, clearly indicate the presence of a 
bright HII region at the supernova location. It is hence quite likely that the 
object detected by \cite{itagaki09} corresponds to the underlying HII region.

\section{Summary}
We present in this paper optical photometry and medium resolution optical 
spectroscopy of the type Ib supernova SN 2009jf, spanning a period from 
$\sim 15$ days before $B$ band maximum to $\sim 250$ days post maximum. 
SN 2009jf reached a $B$ maximum on JD 2455119.46, with an absolute magnitude 
$M_{B} = -17.58\pm0.19$ magnitude. A slow post-maximum decline is indicated by
the broad light curves.  The peak bolometric flux implies $\sim 0.17$ M$_\odot$ 
of $^{56}$Ni was synthesized during the explosion.
 
The spectral evolution of SN 2009jf is typical of type Ib class, but with an 
early emergence of helium lines.  He\,I 5876~\AA\ is clearly identified in the
first spectrum obtained 15 days before maximum, at a velocity of $\sim 16000$ 
km sec$^{-1}$. This early emergence of helium lines is likely due to a substantial
mixing of the inner layers of the ejecta. The [O\,I] 6300-6364~\AA\ line seen in
the nebular spectrum is multi-peaked and asymmetric, with a sharp, stronger blue
peak. This is explained by the complex ejecta structure of an aspherical 
explosion. The absolute flux of this line indicates the mass of oxygen ejected
during the explosion to be $\ga 1.34$ M$_\odot$. 

A qualitative analysis of the light curve and spectra of SN 2009jf indicates 
that SN 2009jf is an energetic explosion of a massive star. The mass of the 
ejecta and kinetic energy of explosion are estimated to be 
M$_{\rm ej}$= $4-9$ M$_\odot$ and K$_{\rm E}$ = $3-8$ $\times 10^{51}$ erg,
respectively. The main sequence mass of the progenitor star is estimated to be  
$\ga 20- 25$ M$_\odot$.  
  
\section*{Acknowledgements}

We would like to thank the anonymous referee for the constructive comments,
which helped in improving the paper.
We thank all the observers of the 2-m HCT who kindly provided part of their
observing time for the supernova observations. This work has been supported 
in part by the DST-JSPS grant DST/INT/JSPS/P-93/2010, the Grant-in-Aid for
Scientific Research of JSPS (20540226) and MEXT (19047004, 22012003)
and by World Premier International Research Center Initiative, MEXT,
Japan.
This work has made use of the NASA Astrophysics Data System and the NASA/IPAC
Extragalactic Database (NED) which is operated by Jet Propulsion Laboratory,
California Institute of Technology, under contract with the National
Aeronautics and Space Administration.


\begin{thebibliography}{99}
\bibitem[\protect\citeauthoryear {Anupama et al.}{2005}]{anupama05} Anupama, G.C., Sahu, D.K., Deng, J., Nomoto, K., Tominaga, N., Tanaka, M., Mazzali, P.A., Prabhu, T.P., 2005, ApJ, 631, L125
\bibitem[\protect\citeauthoryear {Arnett}{1982}]{arnett82} Arnett, W.D., 1982, ApJ, 253, 785
\bibitem[\protect\citeauthoryear {Begelman \& Sarazin}{1986}]{begelman86}Begelman, M.C. \& Sarazin, C.L., 1986, ApJ, 302, L59
\bibitem[\protect\citeauthoryear {Berger \& Soderberg}{2008}]{berger08}Berger, E. \& Soderberg A.M., 2008, GCN, 7159
\bibitem[\protect\citeauthoryear {Bessell, Castelli \& Plez}{1998}]{bessel98} Bessell M.S., Castelli F., Plez B.,  1998, A\&A, 333, 231.
\bibitem[\protect\citeauthoryear {Branch et al.}{2002}]{branch02}Branch, D., Benetti, S., Kasen, D., Baron, E., Jeffery, D.J., Hatano, K., Stathakis, R.A., Filippenko, A.V. et al. 2002, ApJ, 566, 1005

\bibitem[\protect\citeauthoryear {Deng et al.}{2000}]{deng00}Deng, J.S., Qiu, Y.L., Hu, J.Y., Hatano, K., Branch, D. 2000, ApJ, 540, 452
\bibitem[\protect\citeauthoryear {Elmhamdi et al.}{2004}]{elmhamdi04}Elmhamdi, A., Danziger, I.J., Cappellaro, E., Della Valle, M., Gouiffes, C., Phillips, M.M., and Turatto, M. 2004, A\&A, 426, 963
\bibitem[\protect\citeauthoryear {Elmhamdi et al.}{2006}]{elmhamdi06}Elmhamdi, A., Danziger, I.J., Branch, D., Leibundgut, B., Baron, E., Kirshner, R.P., 2006, A\&A, 450, 305

\bibitem[\protect\citeauthoryear {Ensman \& Woosley}{1988}]{ensman88}Ensman, L.M. \& Woosley, S.E.  1988, ApJ, 333, 754 
\bibitem[\protect\citeauthoryear {Filippenko et al.}{1990}]{filipenko90}Filippenko, A.V., Porter, A.C., Sargent, W.L.W. 1990, AJ, 100, 1575
\bibitem[\protect\citeauthoryear {Fransson \& Chevalier}{1987}]{fransson87}Fransson, C.,  \& Chevalier, R.  1987, ApJ, 322, L15 
\bibitem[\protect\citeauthoryear {Fransson \& Chevalier}{1989}]{fransson89}Fransson, C.,  \& Chevalier, R.  1989, ApJ, 343, 323 
\bibitem[\protect\citeauthoryear {Galama et al.}{1998}]{galama98} Galama, T.J., Vreeswijk, Paradijs, J., Kouveliotou, C., Augusteijn, T., Bohnhardt, H., Brewer, J.P., Doublier, V. et al., 1998, Nature, 396, 670
\bibitem[\protect\citeauthoryear {Gomez \& Lopez}{1994}]{gomez94}Gomez, G. \& Lopez, R., 1994, AJ, 108, 195
\bibitem[\protect\citeauthoryear {Hamuy et al.}{2002}]{hamuy02} Hamuy, M., Jose, M., Pinto, P.A., Phillips, M.M., Suntzeff, N.B., Blum, R.D., Oslen, K.A.G., Pinfield, D.J.  et al.  2002 AJ, 124, 417
\bibitem[\protect\citeauthoryear {Hamuy et al.}{2009}]{hamuy09}Hamuy, M., Deng, J., Mazzali, P.A., Morrell, N.I., Phillips, M.M., Roth, M., Gonzlez, S., Thomas-Osip, J. et al. 2009, ApJ, 703, 1612\
\bibitem[\protect\citeauthoryear {Harkenss et al.}{1987}]{harkenss87} Harkenss, R.P., Wheeler, J.C., Margon, B., Downes, R.A., Kirshner, R.P., Uomoto, A., Barker, E.S., Cochran, A.L. et al. 1987, ApJ, 317, 355
\bibitem[\protect\citeauthoryear {Horne}{1986}]{horne86}Horne, K., 1986, PASP, 98, 609 
\bibitem[\protect\citeauthoryear {Itagaki, Kaneda \&  Yamaoka}{2009}]{itagaki09}Itagaki, K., Kaneda, H., Yamaoka, H., 2009,  CBET 1955 
\bibitem[\protect\citeauthoryear {Jester et al.}{2005}]{jester05}Jester, S., Schneider, D.P., Richards, G., T., Green, R.F., Schmidt, M. Hall, P.B., Strauss, M.A. et al. 2005, AJ, 130, 873
\bibitem[\protect\citeauthoryear {Kasliwal et al.}{2009}]{kasliwal09} Kasliwal, M.M., Howell, J.L., Fox, D., Quimby, R., Gal-Yam, A. 2009 ATel 2218
\bibitem[\protect\citeauthoryear {Krisciunas \& Rest}{2000}]{krisciunas00}Krisciunas, K. \& Rest, A.,  2000, IAUC 7382
\bibitem[\protect\citeauthoryear {Landolt}{1992}] {landolt92} Landolt A.U., 1992, AJ, 104, 340

\bibitem[\protect\citeauthoryear {Leonard et al.}{2006}]{leonard06}Leonard, D.C., Filippenko, A.V., Ganeshalingam, M., Serduke,F.J.D., Li, W., Swift, B.J., Gal-Yam, A., Foley, R.J.  et al. 2006, Nature, 440, 505
\bibitem[\protect\citeauthoryear {Li}{2007}]{li07}Li, L. -X., 2007, MNRAS, 375, 240
\bibitem[\protect\citeauthoryear {Li, Cenko \& Filippenko}{2009}]{li09} Li, W., Cenko, S.B., Filippenko, A.V., 2009, CBET 1952
\bibitem[\protect\citeauthoryear {Lucy}{1991}]{lucy91}Lucy, L.B., 1991, ApJ, 383, 308

\bibitem[\protect\citeauthoryear {Maeda et al.}{2002}]{maeda02}Maeda, K., Nakamura, T.,  Nomoto, K., Mazzali, P.A., Patat, F., Hachisu, I.   2002, ApJ, 565, 405
\bibitem[\protect\citeauthoryear {Maeda et al.}{2006}]{maeda06} Maeda, K., Mazzali, P.A., Nomoto, K. 2006, ApJ, 645, 1331 
\bibitem[\protect\citeauthoryear {Maeda et al.}{2007}]{maeda07} Maeda, K., Tanaka, M., Nomoto, K., Tominaga, N., Kawabata, K., Mazzali, P.A., Umeda, H., Suzuki, T., Hattori, T. 2007, ApJ, 666, 1069 
\bibitem[\protect\citeauthoryear {Martin et al.}{1999}]{martin99} Martin, R., Williams, A., Woodings, S., Biggs, J. Verveer, A., 1999, IAUC, 7310
\bibitem[\protect\citeauthoryear {Matheson et al.}{2001}]{matheson01}Matheson, T., Filippenko, A.V., Li, W., Leonard, D.C., 2001, AJ, 121, 1648
\bibitem[\protect\citeauthoryear {Maurer et al.}{2010}]{maurer10} Maurer, I., Mazzali, P.A., Taubenberger, S., Hachinger, S., 2010, arXiv:1007.1881
\bibitem[\protect\citeauthoryear {Maza \& Ruiz}{1989}]{maza89} Maza, J.  \& Ruiz, M.T.,  1989, ApJS, 69, 353
\bibitem[\protect\citeauthoryear {Mazzali et al.}{2001}]{mazzali01}Mazzali, P.A., Nomoto, K., Patat, F., Maeda K. 2001, ApJ, 559, 1047
\bibitem[\protect\citeauthoryear {Mazzali et al.}{2005}]{mazzali05}Mazzali, P.A., Kawabata, K.S., Maeda K., Nomoto, K., Filippenko, A.V., Ruiz, R., Benetti, S., Pian, E. et al. 2005, Science, 308, 1284 

\bibitem[\protect\citeauthoryear {Mazzali et al.}{2008}]{mazzali08} Mazzali, P.A., Valenti, S., Della Valle, M., Chincarini, G., Sauer, D.N., Benetti, S., Elena, P., Piran, T.   et al. 2008, Science, 321, 1185
\bibitem[\protect\citeauthoryear {Mazzali et al.}{2009}]{mazzali09}Mazzali, P.A., Deng, J., Hamuy, M., Nomoto, K., 2009, ApJ, 703, 1624
\bibitem[\protect\citeauthoryear {Mazzali et al.}{2010}]{mazzali10}Mazzali, P.A., Maurer, I., Valenti, S., Kotak, R., Hunter, D. 2010, MNRAS, 408, 87
\bibitem[\protect\citeauthoryear {Modjaz et al.}{2009}]{modjaz09} Modjaz, M., Li, W., Butler, N., Chornock, R., Perley, D., Blondin, S., Bloom, J. S., Filippenko, A. V., Kirshner, R. P et al. 2009, ApJ, 702, 226.
\bibitem[\protect\citeauthoryear {Nadyozhin}{1994}]{nadyozhin94} Nadyozhin, D.K. 1994, ApJS, 92, 527
\bibitem[\protect\citeauthoryear {Nomoto et al.}{1994}]{nomoto94} Nomoto, K., Yamaoka, H., Pols, O.R. et al. 1994, Nature, 371, 227.
\bibitem[\protect\citeauthoryear {Nomoto et al.}{2003}]{nomoto03} Nomoto, K., Maeda,
K., Umeda, H., et al. 2003, IAU Symp. 212, A Massive Star Odyssey: from Main
Sequence to Supernovae, ed. K.A. Van der Hucht et al. (San Francisco: ASP), 395
(astro-ph/0209064)

\bibitem[\protect\citeauthoryear {Nomoto et al.}{2004}]{nomoto04} Nomoto, K., Maeda,
K., Mazzali, P.A., et al. 2004, Astrophysics and Space Science Library 302, Stellar
Collapse, ed. C.L. Fryer (Dordrecht: Kluer), 277 (astro-ph/0308136)

\bibitem[\protect\citeauthoryear {Nomoto et al.}{2006}]{nomoto06} Nomoto, K.,
Tominaga, N., Umeda, H., et al. 2006, Nuclear Physics A, 777, 424 (astro-ph/0605725)
\bibitem[\protect\citeauthoryear {Nugent et al.}{1995}]{nugent95} Nugent, P., Branch, D., Baron, E., Fisher, A., Vaughan, T., 1995, Physical Review Letters, 75, 394

\bibitem[\protect\citeauthoryear {Patat et al.}{2001}]{patat01}Patat, F., Cappellaro, E., Danziger, J., Mazzali, P.A., Sollerman, J., Augusteijn, T., Brewer, J., Doublier, V. et al. 2001, ApJ, 555, 900
\bibitem[\protect\citeauthoryear {Podsiadlowski et al.}{2004}]{pods04} Podsiadlowski, P., Langer, N., Poelarends, A. J. T., Rappaport, S., Heger, A., \& Pfahl, E. 2004, ApJ, 612, 1044
\bibitem[\protect\citeauthoryear {Quimby et al.}{2007}]{quimby07}Quimby, R.M., Aldering, G., Wheeler, J.C., Hoflich, P., Akerlof, C.W., Rykoff, E.S. 2007, ApJ, 668, L99

\bibitem[\protect\citeauthoryear {Richardson, Branch \& Baron}{2006}]{richardson06}Richardson, D., Branch, D., Baron, E. 2006, AJ, 131, 2233.
\bibitem[\protect\citeauthoryear {Richmond et al.}{1996}]{richmond96}Richmond, M.W., Treffers, R.R., Filippenko, A.V., Paik, Y., Leibundgut, B., Schulman, E. \& Cox, C.V. 1996, AJ, 111, 327
\bibitem[\protect\citeauthoryear {Sahu, Anupama \&  Gurugubelli}{2009}]{sahu09a}Sahu, D.K., Anupama, G.C., Gurugubelli, U.K., 2009, CBET 1955.
\bibitem[\protect\citeauthoryear {Sahu et al.}{2009}]{sahu09b}Sahu, D.K., Tanaka, M., Anupama, G.C., Gurugubelli, U.K., Nomoto, K., 2009, ApJ, 697, 676
\bibitem[\protect\citeauthoryear{Schlegel \& Krishner}{1989}]{schlegel89} Schlegel, E.M. \& Krishner, R.P.  1989, AJ, 98, 577
\bibitem[\protect\citeauthoryear {Schlegel, Finkbeiner \& Davis}{1998}] {schlegel98}  Schlegel D.J., Finkbeiner D.P., Davis M., 1998, ApJ, 500, 525

\bibitem[\protect\citeauthoryear {Silverman et al.}{2009}]{silverman09}Silverman, J.M., Mazzali, P., Chornock, R., Filippenko, A.V., Clocchiatti, A., Phillips, M.M., Ganeshlingam, M., Foley, R.J., 2009, PASP, 121, 689
\bibitem[\protect\citeauthoryear {Soderberg et al.}{2008}]{soderberg08}Soderberg, A.M., Berger, E., Page, K.L., Schady, P., Parrent, J., Pooley, D., Wang, X.-Y., Ofek, E.O. et al. 2008, Nature, 453, 469
\bibitem[\protect\citeauthoryear {Sollerman, Leibundgut \& Spyromilio}{1998}]{sollerman98}Sollerman, J., Leibundgut, B., Spyromilio, J. 1998, A\&A, 337, 207
\bibitem[\protect\citeauthoryear {Stalin et al.}{2008}]{stalin08}Stalin, C.S., Hegde, M., Sahu, D.K., Parihar, P.S., Anupama, G.C., Bhatt, B.C., Prabhu, T.P.  2008, BASI, 36, 111
\bibitem[\protect\citeauthoryear {Stritzinger et al.}{2002}]{strit02}Stritzinger, M., Hamuy, M.,  Suntzeff, N. B., Smith, R. C., Phillips, M. M., Maza, J.; Strolger, L.-G., Antezana, R. et al. 2002, AJ, 124, 2100
\bibitem[\protect\citeauthoryear {Stritzinger et al.}{2009}]{strit09}Stritzinger, M.,  Mazzali, P., Phillips, M. M., Immler, S., Soderberg, A., Sollerman, J.,  Boldt, L., Braithwaite, J. et al. 2009, ApJ, 696, 713.
\bibitem[\protect\citeauthoryear {Tanaka et al.}{2007}]{tanaka07}Tanaka, M., Maeda, K., Mazalli, P.A., Nomoto, K., 2007, ApJ, 668, L19
\bibitem[\protect\citeauthoryear {Tanaka et al.}{2009}]{tanaka09} Tanaka, M., Tominaga, N., Nomoto, K., Valenti, S., Sahu, D.K., Minezaki, T., Yoshii, Y., Yoshida, M., Anupama, G.C., et al. 2009, ApJ, 692, 1131
\bibitem[\protect\citeauthoryear {Taubenberger et al.}{2009}]{taubenberger09} Taubenberger, S., Valenti, S., Benetti, S., Cappellaro, E., Della Valle, M., Elisa-Rosa, N., Hachinger, S., Hillebrandt, W., Maeda, K.  et al. 2009, MNRAS, 397, 677
\bibitem[\protect\citeauthoryear {Uomoto}{1986}]{uomoto86}Uomoto, A., 1986, ApJ, 310, L35 
\bibitem[\protect\citeauthoryear {Wang et al.}{2003}]{wang03}Wang, L.,Baade, D., Hoflich, P., Wheeler, J.C.  2003, ApJ, 592, 457
\bibitem[\protect\citeauthoryear {Woosley, Langer \& Weaver}{1993}]{woosley93} Woosley, S.E., Langer, N., Weaver, T.A., 1993, ApJ, 411, 823.
\end{thebibliography}
\end{document}